\pdfoutput=1
\documentclass[useAMS,usenatbib]{mn2e}
\usepackage{amsmath, float, bm}
\usepackage{txfonts}
\usepackage{fleqn}   
\usepackage{graphicx}

\newcommand{\be}{\begin{equation}}
\newcommand{\ee}{\end{equation}}
\newcommand{\bdm}{\begin{displaymath}}
\newcommand{\edm}{\end{displaymath}}

\newcommand{\bea}{\begin{eqnarray}}
\newcommand{\eea}{\end{eqnarray}}
\newcommand{\ba}{\begin{align}}
\newcommand{\ea}{\end{align}}

\title[Constraints on particle acceleration sites in the Crab Nebula]
{Constraints on particle acceleration sites in the Crab Nebula from relativistic MHD simulations}
\author[B. Olmi et al.]{B. Olmi$^{1,2,3}$\thanks{E-mail:
barbara.olmi@unifi.it}, L. Del Zanna$^{1,2,3}$, E. Amato$^{2}$, N. Bucciantini$^{2,3}$\\
$^{1}$Dipartimento di Fisica e Astronomia, Universit\`a degli Studi di Firenze, Via G. Sansone 1, 50019 Sesto F.~no  (Firenze), Italy\\
$^{2}$INAF - Osservatorio Astrofisico di Arcetri, Largo E. Fermi 5, 50125 Firenze, Italy\\
$^{3}$INFN - Sezione di Firenze, Via G. Sansone 1, 50019 Sesto F.~no  (Firenze), Italy}

\begin{document}
 
\date{Accepted/Received}

\maketitle

\label{firstpage}

\begin{abstract}
The Crab Nebula is one of the most efficient accelerators in the Galaxy and the only galactic source showing direct evidence of PeV particles. In spite of this, the physical process behind such effective acceleration is still a deep mystery. While particle acceleration, at least at the highest energies, is commonly thought to occur at the pulsar wind termination shock, the properties of the upstream flow are thought to be non-uniform along the shock surface, and important constraints on the mechanism at work come from exact knowledge of where along this surface particles are being accelerated. Here we use axisymmetric relativistic MHD simulations to obtain constraints on the acceleration site(s) of particles of different energies in the Crab Nebula. Various scenarios are considered for the injection of particles responsible for synchrotron radiation in the different frequency bands, radio, optical and X-rays. The resulting emission properties are compared with available data on the multi wavelength time variability of the inner nebula. 
Our main result is that the X-ray emitting particles are accelerated in the equatorial region of the pulsar wind. 
Possible implications on the nature of the acceleration mechanism are discussed.
\end{abstract}


\section{INTRODUCTION}
Pulsars and their relativistic outflows are the most powerful galactic particle accelerators, and non-thermal emission from Pulsar Wind Nebulae (PWNe) is the best available source of information about the physics of these compact objects, including their role as antimatter factories. The prototype of the class of PWNe, and one of the best studied object among all astronomical sources, is certainly the Crab Nebula (CN hereafter). This nebula, characterised by a very well studied, very broad-band, non-thermal spectrum, represents an excellent laboratory for high-energy astrophysics.

The source that powers the nebular emission is the central fast-spinning and magnetized neutron star (a young 33~ms pulsar), that, by means of rotational energy losses, produces a highly magnetized, cold, relativistic wind primarily made of electron-positron pairs. The wind impact on the slowly expanding supernova ejecta induces the formation of a reverse shock, the so-called Termination Shock (TS), that ensures deceleration of the outflow and dissipation of its bulk energy at a distance of a few arcsec from the pulsar. Based on observations of the nebular synchrotron emission, this termination shock is believed to be the place where about 20-30\% of the outflow energy is thought to be converted into accelerated particles up to energies of order 1PeV. 

The study of PWNe, and of the CN in particular, is crucial for many fundamental aspects of high-energy astrophysics, like neutron stars physics, cosmic ray production, and the physics of highly relativistic, collisionless shocks \citep[for a recent review see][]{Amato:2014}. Pulsars are believed to be the most important galactic antimatter factories, thus providing a possible contribution to explain the positron excess recently found by PAMELA and AMS02 \citep{PAMELA-coll.:2009, AMS-02-coll.:2013}. Unfortunately the rate of particle extraction and the details of the subsequent cascading process in the inner magnetosphere are still uncertain, usually quantified with the so-called pair \emph{multiplicity} $\kappa$. Moreover, the observed extremely efficient acceleration of the wind particles at the TS is in contrast with shock acceleration theories, given that ultrarelativistic, transverse shocks (the magnetic field in the wind is believed to be mainly toroidal at large distances from the pulsar) are the worst possible configuration for Fermi-type processes to operate effectively \citep{Arons:2012, Sironi:2013}.

Detailed modelling of the CN dynamics and emission, with a special attention to its inner region and careful comparison with observations, may help clarify the situation. The inner region of the CN has been known to be highly variable at optical wavelengths since the late '60s, when the so-called \emph{wisps} were first observed by \citet{Scargle:1969}. These arc-shaped bright features are periodically produced where the TS is expected to be located, the first bright ring surrounding the central dark zone. These bright arcs of emission then move radially outwards at mildly relativistic velocities, as expected for a post-shock hydrodynamic flow. Due to their apparent connection with the particle acceleration site, it is clear how a precise modelling of such features can provide clues on the physical mechanisms at work.

In the last decade, thanks to the unprecedented imaging power of \emph{Chandra}, high resolution observations of the inner region of the CN have become available also in the X-ray band, revealing a striking \emph{jet-torus} structure, knots and rings \citep{Weisskopf:2000}, previously just barely inferred \citep{Brinkmann:1985, Hester:1995}. These observations prompted a theoretical effort at modelling PWNe, and the CN in particular, within the framework of time-dependent multi-D MHD. Axisymmetric relativistic MHD simulations of the interaction of the pulsar wind with the supernova ejecta have proven to be an excellent tool to investigate the physics of PWNe: not only the overall jet-torus structure, but even very fine details as X-ray knot and rings have been successfully reproduced \citep{Komissarov:2004, Del-Zanna:2004, Del-Zanna:2006}.

MHD simulations also provided an explanation for the CN moving wisps. After optical discovery, the latter have been studied with increasingly high accuracy at several wavelengths, from radio to X-rays \citep{Tanvir:1997, Bietenholz:2001, Hester:2002, Bietenholz:2004, Hester:2008, Schweizer:2013}. A few different interpretations have been proposed in the literature for their phenomenology. \citet{Gallant:1994} and \citet{Spitkovsky:2004} proposed their association with ion-cyclotron waves produced by the presence of relativistic ions in the pulsar wind, while \citet{Begelman:1999} proposed to interpret them as Kelvin-Helmoltz instabilities of shear flows around the TS, though \citet{Bucciantini:2006} later showed that in the magnetized case this process is unlikely to play a role. In reality, the wisps find a very natural interpretation in terms of the properties of the MHD flow around the termination shock \citep{Volpi:2008, Camus:2009}: the periodical ($P\sim 1-3$ years) creation of fluid vortices around the TS, due to magnetosonic oscillations occurring on about a light-crossing time of the shock, is seen to well reproduce outward moving wisps with the observed velocities of $0.3-0.5~c$. The wisps would then simply be Doppler boosted features in a turbulent and highly magnetized environment. This scenario has not basically changed in the aftermath of the first full three-dimensional MHD simulation of the CN \citep{Porth:2014}, where wisps are seen to survive an even more complex flow structure, at least within the stringent limits on both resolution and length of these first simulations, imposed by the much more severe computational costs.

In spite of the successes obtained by MHD models, some questions are still unanswered. One of these, for instance, is the origin of the electrons and positrons responsible for the CN synchrotron radio emission (\emph{radio particles} hereafter). These might have been injected in the nebula at very early stages, and then reaccelerated by local turbulence \citep{Olmi:2014}, rather than being continuously injected at the TS, as is instead the case for the higher energy pairs (\emph{X-ray particles}). The overall synthetic spectrum from one-zone models of the CN is in fact best reproduced if two distinct particle populations are assumed, with a spectral break around the infrared band \citep{Atoyan:1996, Meyer:2010}. 
From a theoretical point of view the two scenarios are very different: if radio particles are injected as part of the outflow we expect a much higher pair multiplicity $\kappa$, and a lower wind Lorentz factor $\gamma$, \citep{Bucciantini:2011} than magnetospheric models are able to predict \citep{Hibschman:2001, Harding:2011, Timokhin:2013}, while the most commonly accepted values of $\kappa\sim 10^4$ and $\gamma\sim 10^6$ \citep{Kennel:1984} are those pertaining two models in which the radio electrons are a relic population of early evolutionary stages.

In order to definitely rule out one of these contrasting hypotheses, in \citet{Olmi:2014} we addressed the question of radio emission morphology and integrated spectrum of the CN, for the first time by means of (two-dimensional) MHD simulations. Using the model parameters that were known to best match the CN morphology, dynamics, and emission, we tested three scenarios for the radio particles. However, due to their long cooling time, we were not able to discriminate between a steady and uniform distribution or a continuous injection as for the X-ray particles: surface brightness radio maps and characterisation of wisps motion both well agree with observations \citep{Bandiera:2002, Bietenholz:2004}, but the two cases are undistinguishable one from the other. The only option we could exclude is that of a relic population of low energy particles which are injected at early times ($\sim 100$ years) and then just advected by the nebular flow. In this case the radio emission does not match the observations at all and we then concluded that, if particles are not injected as part of the wind population, some form of particle re-acceleration (by the turbulent flow or distributed reconnection events) must be continuously at work.

Another important issue to be addressed is the precise location of the acceleration of relativistic particles around the TS. An answer to this question would be particularly important in discriminating sites where shock acceleration can be more efficient. The two mechanisms that are most commonly invoked as the origin of particle acceleration at the CN TS are the first order Fermi process (the most commonly invoked acceleration mechanism in astrophysics) and driven magnetic reconnection of the striped wind, presently the best candidate to explain the particle acceleration needed to reproduce the Crab $\gamma$-ray flares above 100~MeV \citep{Cerutti:2014}. Both processes require special conditions to be at work, which are not realised everywhere along the shock surface. In order for Fermi mechanism to be operative, the flow magnetisation, parametrised by the ratio of Poynting flux to particle kinetic energy flux in the wind, $\sigma$, must satisfy: $\sigma<0.001$ \citep{Spitkovsky:2008, Sironi:2011}. MHD simulations taught us that, if the emission from Crab is to be reproduced, this condition can only be realised in a thin latitude strip around the pulsar equator or in the vicinities of the polar axis, while at intermediate latitudes the flow magnetisation must be substantially larger than this, and likely $\sigma$ of order a few \citep{Komissarov:2013a}. As far as magnetic reconnection is concerned, this can only operate in regions where the wind is striped \citep{Coroniti:1990, Lyubarsky:2001, Kirk:2004, Sironi:2011}, or where there is an O-point dissipation: again these conditions are only realised around the equator and close to the polar axis respectively. 
In addiction for magnetic reconnection to give as broad a spectrum as observed, the pulsar multiplicity must be extremely high, much higher than theory can account for, and even so high that the wind would reconnect before the TS \citep{Lyubarsky:2003}.
The two mechanisms are also very different in terms of the particle spectrum they produce, with Fermi mechanism leading to $N(E)\propto E^{-p}$ with $p \approx 2$, appropriate to account for the energy distribution of the highest energy particles, and magnetic reconnection leading to $1\lesssim p\lesssim 1.5$, appropriate to account for the radio particle spectrum.

One way to test the different scenarios is by comparison with observations of the variability they entail in the inner nebula at various frequencies: since the wisps are seen to start so close to the termination shock (or the location where it is supposed to be), at least at optical and X-ray frequencies, where radiative losses are important, they trace freshly injected particles, and in the simulated maps, their appearance and motion depends on the location at which the emitting particles are injected in the nebula.

In this article we use axisymmetric MHD numerical models of the CN to infer constraints on the acceleration site(s) of the emitting particles. We consider different scenarios for the injection of particles of different energies and compare the resulting simulated wisps with available data at radio, optical and X-ray frequencies. 

The motivation for using these features as probes for the particle acceleration site comes from the fact that wisps observed at various wavelengths, namely in radio, optical, and X-rays, are not coincident features and are seen to propagate at different speeds \citep{Bietenholz:2004, Schweizer:2013}. While the observed discrepancies between radio and optical wisps led \citet{Bietenholz:2004} to conclude that the two populations must have a different acceleration mechanism and/or site, in \citet{Olmi:2014} we showed that wisps are well reproduced without invoking \emph{ad hoc} mechanisms, with the bulk flow acting as the main driver for the observed wisps appearance and motions. However, the issue of the different behaviour at various wavelengths was not addressed in detail and this is the main goal of the present work. Assuming the emitting particles are all accelerated at the TS, including the radio component (thus with a common mechanism), the discrepancies can only be explained by choosing different acceleration regions along the TS for distinct distributions. If particles with different energies are injected at different angles, the paths induced by the post-shock flow structures, and the adiabatic and synchrotron losses, will be also different, thus we do not expect to see identical moving features at all frequencies, as observed.

The paper is structured as follows. In section~\ref{sec:sim_det} we briefly summarise our wind model, provide the definition of the model parameters, and recall the basic methods for obtaining synchrotron emission spectra and maps from MHD simulations. In section~\ref{sec:data_analysis} the details of synthetic data analysis of the simulated wisps are provided. Section~\ref{sec:res1} is devoted to the results in the different scenarios, and section~\ref{sec:res2} to the characterisation of wisps. In section~\ref{sec:concl} the conclusions are drawn.

\section{SIMULATION DETAILS}
\label{sec:sim_det}

\subsection{Wind model and numerical setup}

The MHD axisymmetric simulations presented here are performed by continuously injecting the cold, relativistic pulsar wind from the inner radial boundary (we adopt spherical coordinates $r$ and $\theta$). The final morphology and characteristics of the simulated PWN are strongly dependent on the chosen pulsar wind model, here we thus provide all details. The energy flux is defined roughly following the \textit{split-monopole} solution \citep{Michel:1973}, which predicts $F(r,\theta) \propto r^{-2} \sin^2\theta$ at large distances from the pulsar. The exact expression of the energy flux employed in our simulation is
\begin{equation}
\label{eq:e_flux-1}
 F(r,\theta) \equiv c ( n \, m_\mathrm{e} c^2 \gamma_0^2 +  B^2/4\upi ) = \frac{L_0 }{4 \upi r^2} \mathcal{F}(\theta)\,  , 
\end{equation}
where we assume a purely radial flow with isotropic Lorentz factor $\gamma_0$. $B \equiv B_\phi$ is the embedded purely toroidal magnetic field, $n$ the particle number density of the pulsar outflow, $L_0$ the pulsar spin-down luminosity (taken as constant in time) and $m_\mathrm{e}$ the electron mass. Moreover, we choose
\begin{equation}
\label{eq:e_flux-2}
\mathcal{F}(\theta)=\frac{\alpha + (1-\alpha)\sin^2\theta}{1-(1-\alpha)/3}\,,
\end{equation}
where we define $\alpha \ll 1$ as the \emph{wind anisotropy parameter}, basically the ratio of polar to equatorial energy fluxes. 

The magnetic field is defined as
\begin{equation}
\label{eq:B_field-1}
B(r,\theta)= \sqrt{\frac{\sigma_0 L_0}{c}}\frac{\mathcal{G}(\theta)}{r}\,,
\end{equation}
where $\sigma_0$ is the \emph{wind magnetisation parameter}. The angular dependence of the field is described by the function
\begin{equation}
\label{eq:B_field-2}
\mathcal{G}(\theta)=\sin\theta \tanh \left[ b\left(\frac{\upi}{2}-\theta \right)\right]\,,
\end{equation}
that takes into account both the split-monopole latitude dependence and the possibility that the striped part of the wind may dissipate. Indeed, from the striped wind model, it is expected that the pulsar outflow is characterised by a belt around the pulsar equator with alternating zones of different field polarities, where magnetic dissipation phenomena take place reducing the magnetization. The parameter $b$ in \eqref{eq:B_field-2} is exactly related to the width of this belt: large $b$ values lead to the pure split-monopole scenario (a narrow striped zone), while low values, as $b \sim 1$, give rise to a modulation of the field strength at all latitudes \citep{Del-Zanna:2006}. 

Given the previous assumptions, the wind magnetization depends in our model on the angle alone as
\begin{equation}
\sigma(\theta)=\frac{B^2}{4 \upi n m_\mathrm{e} \gamma_0^2 c^2} = \frac{\sigma_0 \mathcal{G}^2}{\mathcal{F} - \sigma_0 \mathcal{G}^2}\,.
\end{equation}
Finally, the particle numerical density injected in the pulsar wind is 
\begin{equation}
n(r,\theta) = \frac{L_0}{4\upi c^3 m_\mathrm{e} \gamma_0^2} \frac{1}{r^2}
\left[ \mathcal{F}(\theta) - \sigma_0 \mathcal{G}^2(\theta) \right]\, ,
\label{eq:n}
\end{equation}
easily obtained from Eqs.~(\ref{eq:e_flux-1}) and (\ref{eq:B_field-1}).

Numerical simulations are performed with the ECHO code \citep{Del-Zanna:2003,Del-Zanna:2007}. In our models of PWNe we assume, as usual, the equation of state of an ultra-relativistic plasma and $v_\phi = B_r = B_\theta \equiv 0$. The numerical setup is very similar to that described in \citet{Olmi:2014}, here with a lower resolution, namely the one that best balances the computational costs and a sufficient resolution in the inner zone of the nebula. The computational box is made up of 512 cells in the radial direction, with the radius ranging from $r_\mathrm{min} = 0.05$~ly to $r_\mathrm{max} = 10$~ly, and characterised by a logarithmic stretching ($dr \propto r)$, whereas the angular domain ranges from $\theta_\mathrm{min} = 0$ to $\theta_\mathrm{max} = \upi$, and it contains 256 cells. 

As far as the initial conditions are concerned, the inner wind zone is surrounded by a spherically expanding shell of cold, dense supernova ejecta, propagating in an external medium at standard ISM conditions. We choose values of the free parameters that best match the observations, both in terms of nebular morphology and spectral properties \citep{Del-Zanna:2006, Olmi:2014}, summarised in Tab.~\ref{tab:init}. 

\begin{table}
\centering
\begin{tabular}{ccccc}
\hline
Wind     &   $L_0=5\cdot10^{38}$ erg s$^{-1}$ & $\gamma_0=100$  & $\alpha=0.1$ & $\sigma_0=0.025$   \\
Ejecta     &  $M_{\mathrm{ej}}= 6 M_\odot$  & $E_{\mathrm{ej}} = 10^{51}$ erg & $\varv\propto r$   \\
ISM     &     $n= 1$ cm$^{-3}$  &  $T=10^4$~K &  & \\
 \hline
\end{tabular}
 \caption{Values of the parameters used in the simulations.}
\label{tab:init}
\end{table}

\subsection{Calculation of the synchrotron emission}
\label{sec:sync}
As in our previous analyses \citep{Volpi:2008, Olmi:2014}, we consider two distinct families of emitting particles: those responsible for the radio emission and those emitting in the X-rays. 
Usually optical emitting particles are taken as part of the same population to which X-ray emitting ones belong. However, 
multifrequency campaigns have shown that wisps not only are different at radio and optical wavelengths \citep{Bietenholz:2004}, but they also differ at optical and X-ray frequencies \citep{Schweizer:2013}. As we will see in the following, this observation can only be accounted for if the injection sites of optical and X-ray emitting particles are not perfectly coincident. A natural assumption, to have wisps that are not coincident in any two of the three observational wavebands, seems that of having radio and X-ray emitting particles injected in different sites and the optical emission produced as the sum of the contribution of radio and X-ray emitting particles.

The spectral shape that we adopt in this work is slightly more complex than that used by \citet{Olmi:2014}, with the introduction of a high energy exponential cut-off for both power-laws. Exponential cut-offs were already used by \citet{Volpi:2008} and had been replaced by sharper falls of the distribution function in our previous work \citep{Olmi:2014} for the sake of simplicity. We are now forced to reintroduce them in an attempt at interpreting the optical emission as partly contributed by the low energy (radio emitting) particles. In fact, a shallower (only exponential) cut-off is crucial if one wants a substantial contribution at optical frequencies of both families of particles without overproducing the infrared emission.

The radio and X-ray emitting particles are injected with spectra respectively given by:
\be
f_{0\mathrm{R}}(\epsilon_0)\propto \left\{
\begin{array}{lcl}
0 & {\rm if} & \epsilon_0 < \epsilon_{\rm minR}, \\
\epsilon_0^{-p_\mathrm{R}} \exp(-\epsilon_0/\epsilon_\mathrm{R}^*) &  {\rm if} & \epsilon_0 > \epsilon_{\rm minR},\\
\end{array}
\right.
\ee
 \be
 f_{0\mathrm{X}}(\epsilon_0)\propto \left\{ 
 \begin{array}{lcl}
 0 & {\rm if} & \epsilon_0 < \epsilon_{\rm minX}, \\
\epsilon_0^{-p_\mathrm{X}} \exp(-\epsilon_0/\epsilon_\mathrm{X}^*) &  {\rm if} & \epsilon_0 > \epsilon_{\rm minX}, \\
\end{array}
\right.
 \label{eq:Xpart}
 \ee
 where $\epsilon_0$ is the Lorentz factor of the particle at the injection site.
The parameters that appear in this description, namely power-law indices and cut-off energies, are all determined based on comparison of the simulated emission with the data, and in such a way that two requirements are satisfied: 1) the integrated emission spectrum is correctly reproduced; 2) the optical emission is partly contributed by the low and high energy particle populations. The values of the parameters on which the results we discuss in the following are based are: $p_\mathrm{R}=1.6$, $\epsilon_{\rm minR}=10^3$, $\epsilon_\mathrm{R}^*=2 \times10^6$, 
$p_\mathrm{X}=2.8$, $\epsilon_{\rm minX}=1.5 \times 10^6$ and $\epsilon_\mathrm{X}^*=10^{10}$.
 
The evolved distribution functions, for both species $s$,  at any place in the nebula and at any time, are determined by the conservation of particle number along the streamlines and by the adiabatic and synchrotron losses that particles undergo in their bulk motion. We use the expression
\begin{equation}
\label{eq:f_wind}
f_{s}(\epsilon)=\left( \frac{n}{n_0}\right)^{4/3} \left( \frac{\epsilon_0}{\epsilon}\right)^2 f_{0s}(\epsilon_0), \quad 
\epsilon_0=\left(\frac{n_0}{n}\right)^{1/3} \frac{\epsilon}{1-\epsilon/\epsilon_{\infty}}\,,
\end{equation}
where $\epsilon_\infty$ is the maximum attainable particle energy, depending on synchrotron cooling, namely the energy at a given location corresponding to an injection with $\epsilon_0 \rightarrow \infty$. Together with the particle density $n$ and its value just downstream of the wind TS $n_0$, $\epsilon_\infty$ is thus one of the three particle tracers that allow us to produce the nebular synthetic emission. These are evolved in the code, for each species $s$ and for each injection location (see below), according to \citet{Del-Zanna:2006} and \citet{Olmi:2014}.

The local synchrotron emission coefficient for an ultrarelativistic electron (or positron) immersed in the comoving magnetic field $\bm{B}^\prime$, for given observer's direction $\bm{n}^\prime$ and frequency $\nu^\prime$, is
\begin{equation}
	j^\prime_\nu(\nu^\prime,\bm{n}^\prime)=\int \mathcal{P}(\nu^\prime, \epsilon) f(\epsilon) d\epsilon\,,
\end{equation}
where $\mathcal{P}(\nu^\prime, \epsilon) $ is the synchrotron spectral power per unit frequency
\begin{equation}
	\mathcal{P(\nu^\prime,\epsilon)} = \frac{ 3 \sqrt{3} e^4 }{ 4\upi m_\mathrm{e}^2 c^3 } \frac{ {B^\prime_\perp}^2  \epsilon^2 }{ \nu^\prime_\mathrm{c}  } F\left( \frac{\nu^\prime}{\nu^\prime_\mathrm{c}}\right)\,.
\end{equation}
Hereafter the apexed quantities refer to the comoving frame.
In the previous formula $F\left( \nu^\prime / \nu^\prime_\mathrm{c} \right)$ is the synchrotron characteristic function, which peaks at about the critical frequency $\nu^\prime_\mathrm{c}=(3e)/(4\upi m_e c)B^\prime_\perp \epsilon^2$ \citep[see][]{Rybicki:1979}. With $B^\prime_\perp = |\bm{B}^\prime \times \bm{n}^\prime|$ we define the magnetic field component normal to the observer line of sight $\bm{n}$. 
The emissivity in the observer reference frame is then obtained by applying the relevant relativistic transformations \citep[see][]{Del-Zanna:2006}. Here we just remind that the frequency and the emission coefficient transform with the Doppler boosting factor
\begin{equation}
	D = \frac{1}{\gamma \left( 1 - \bbeta \cdot \bm{n} \right)}\,,
\end{equation}
with $\bbeta=\bm{\varv}/c$ and $\gamma$ the Lorentz factor of the fluid. This is responsible for the enlightenment of those features which move towards the observer.

Finally, once the emissivity is known at any point in the nebula, surface brightness maps and total luminosity per unit frequency (basically the Spectral Energy Distribution) are calculated assuming the plasma to be optically thin (as appropriate for the tenuous plasma of PWNe). We have, respectively
\begin{equation}
	I_\nu(x,y) = \int j_\nu(\nu,x,y,z) \mathrm{d}z \,, \quad\quad
	L_\nu = \int \!\! \int I_\nu(x,y,z) \mathrm{d}x \mathrm{d}y \,,
\end{equation}
where $(x,y,z)$ is a Cartesian coordinates system in the observer frame and $z$ defines the line of sight ($\bm{n}$), while $(x,y)$ is the plane of the sky. Integrals are computed over the nebula dimensions.

\section{SYNTHETIC DATA ANALYSIS}
\label{sec:data_analysis}
We have considered several different scenarios, with particles of different energies being injected uniformly along the shock front, mainly around the polar axis, or mainly around the equatorial plane of the pulsar rotation. Various angular amplitudes of the polar and equatorial regions have also been considered.

\begin{figure}
 \includegraphics[scale=0.75]{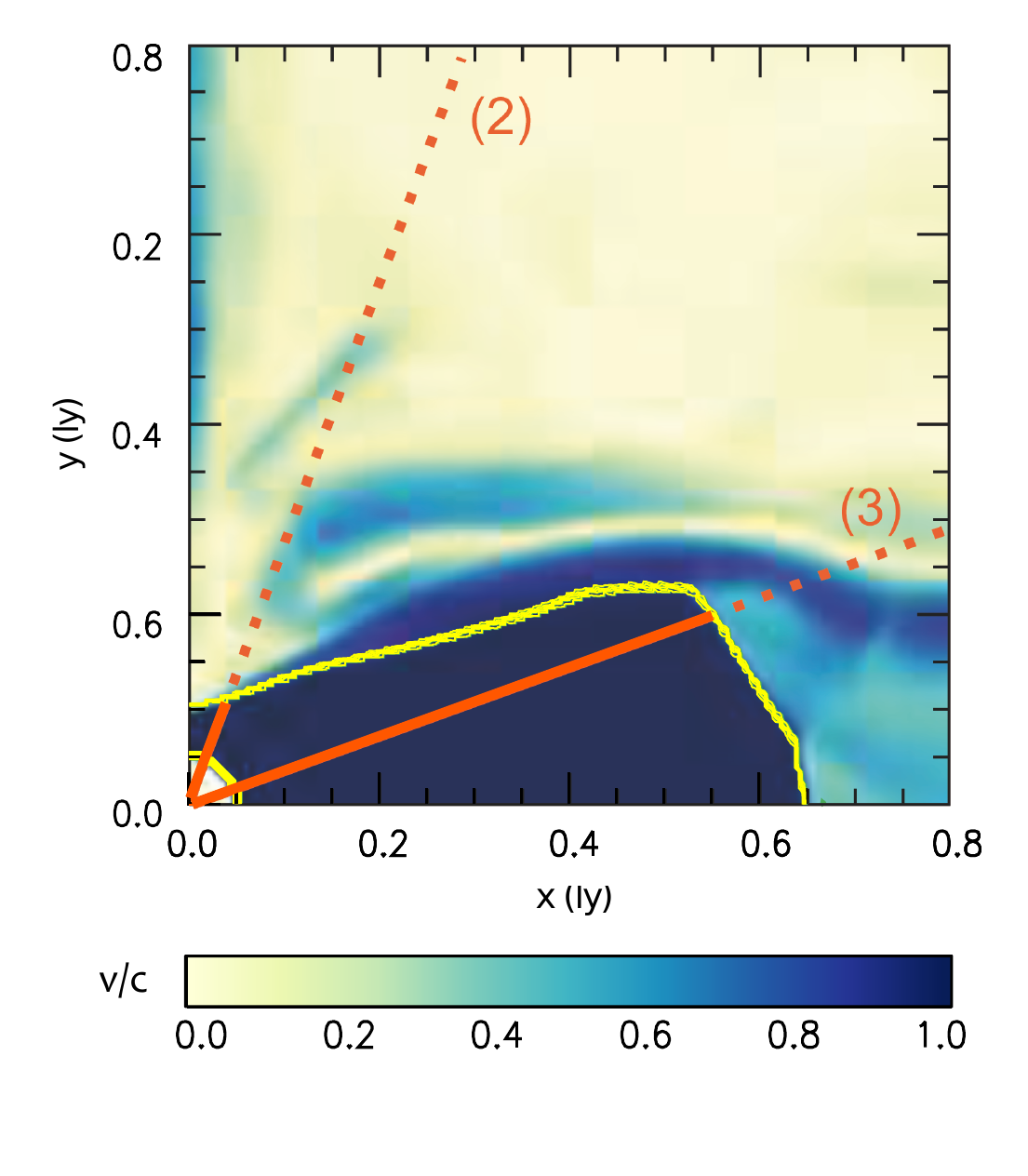}
 \caption{Map of the flow structure, with the inner boundary and the TS highlighted by the yellow solid line. The colours indicate the velocity magnitude in terms of $c$. The two orange solid-dotted lines identify the injection angular regions for case (2) and case (3).}
  \label{fig:TS-zones}
\end{figure}

We list below the different injection geometries. Angular extents are expressed in the upper hemisphere, but symmetry around the equator is implicit. The cases we consider are: 
\begin{enumerate}
 \renewcommand{\theenumi}{(\arabic{enumi})}
	\item uniform injection: particles are injected at all the angles in the nebula between $0\degr$ and $90\degr$;
	\item	wide equatorial region: polar region defined by $\theta \in [0\degr, 20\degr]$ and equatorial one by $\theta \in [20\degr, 90\degr]$;
	\item narrow equatorial region: polar region defined by $\theta \in [0\degr, 70\degr]$ and equatorial one by $\theta \in [70\degr, 90\degr]$.
\end{enumerate}
A few comments on these choices are in order.
Case (2) mimics a scenario in which acceleration is associated with the low-magnetization region close to the polar axis, thanks to the Fermi mechanism, or with O-point dissipation.
Case (3) has been defined in order to roughly match the narrow striped wind region in our reference case with $b=10$ (and $\sigma_0=0.025$). 
In Fig.~\ref{fig:TS-zones} we show the boundaries of the injection regions for these cases, over-plotted on the flow structure around the TS. 

All the choices listed above are considered for all three ranges of particle energies (radio, optical and X-ray emitting particles), namely, particles in each of those energy ranges can be injected in five different regions: wide or narrow polar cap, wide or narrow equatorial belt, and along the entire shock surface. 

As far as the analysis of wisps is concerned, data are produced as a monthly output at the end of the full simulation of the CN evolution, over a period of 10 years between 950 and 960 years \citep[see][for details of the reference simulation]{Olmi:2014}. Using the expressions given in Sec.~\ref{sec:sim_det}, we first calculate the integrated spectra for $t=950$~yr, in order to compute the correct normalisation of the particles spectra for each of the five injection scenarios. Once the normalisations are found, we compute the surface brightness maps, for each case and for all output times, at radio (5~GHz), optical (1~eV) and X-ray (1~keV) frequencies.

\begin{figure}
\includegraphics[scale=.6]{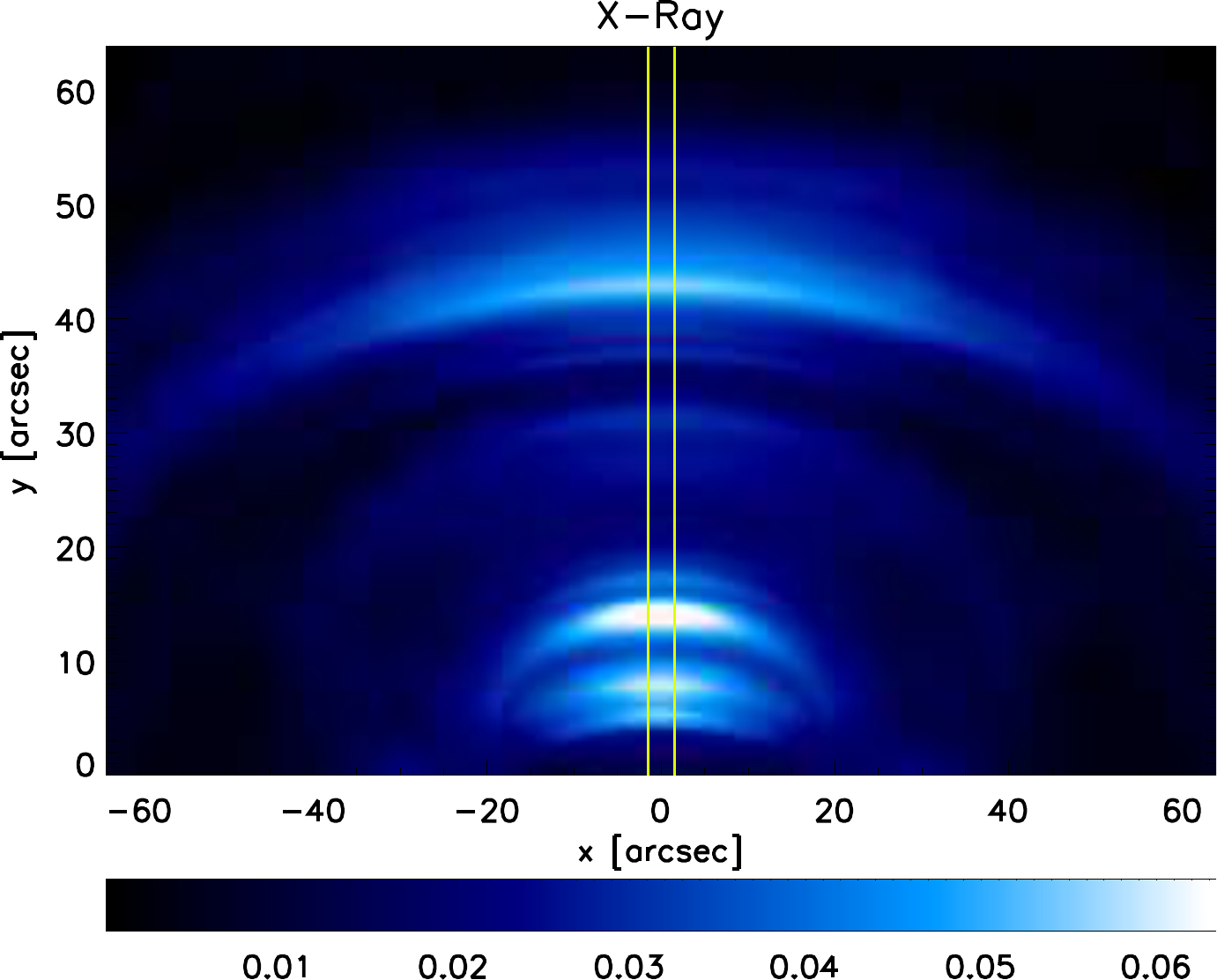}
 \caption{The upper emisphere of the X-ray surface brightness map (1~keV) at $t=950$~yr, in linear scale and in mJy/arcsec$^2$ units. The vertical lines identify the slice used for extraction of the radial profiles.}
  \label{fig:Xwisp}
\end{figure}

As an example, we show in Fig.~\ref{fig:Xwisp} a surface brightness map  at X-ray frequencies.
The emitting particles are assumed to be injected in a narrow equatorial belt (case 3). 
As pointed out in the introduction, the wisps appearance in our simulations is totally due to the combined effects of the locally enhanced magnetic field, just downstream of the TS, and Doppler boosting (channels with significant $\varv/c$ form along the oblique sectors of the shock surface).
 In particular, Doppler boosting is responsible for the angular profile of wisps, as well as for the enhanced brightness of the front side of the nebula with respect to the back side, that appears very faint. The intensity contrast between distinct wisps is, on the contrary, strongly connected to the local magnetic field strength.
The latter is also responsible for the suppression of most of the torus emission (again, we recall that in our axisymmetric simulations the magnetic field in the torus is quite low, $B< 10^{-4}$~G). Concluding, the brightest features are not necessarily the most boosted ones, but surely are those that highlight the magnetic field enhancements in the nebula.

For the analysis of our simulation data, we follow a similar procedure as that applied by \citet{Schweizer:2013} in analysing the observations. The intensity map is cut within a $32\arcsec$ radius from the pulsar and we consider a slice of width $3\arcsec$ in the upper hemisphere, centred on the nebula polar axis (as shown in Fig.~\ref{fig:Xwisp}). For ease of comparison with real data, the intensity is convolved with the appropriate instrumental PSF and only intensity peaks with $I \geq I_\mathrm{max}/3$ are taken into account, where $I_\mathrm{max}$ is the maximum value of the intensity in each map. 
This cut is applied in order to remove the background of weaker variations, that are not useful for the comparison with the data. 
The PSFs employed are those of the instruments used for the observations considered for comparison in each frequency band: in particular we refer to the data analysed in \citet{Schweizer:2013}, \citet{Hester:2002} and \citet{Bietenholz:2004}. The relevant observations were obtained with Chandra for the X-rays (with a FWHM=$0.5\arcsec$), Nord Optical Telescope (NOT) for the optical (FWHM=0.75\arcsec, as determined based on the average seeing of the period of reference), and VLA for the radio frequencies (with a FWHM=$1.8\arcsec$). By using this procedure we can obtain radial profiles for the maximum intensity at every time. 

In the following we show two types of plots to illustrate our results: the radial distance of intensity peaks from the pulsar as a function of time, and, in order to provide information on the relative significance of the different peaks, stack plots of (properly rescaled) brightness profiles, superimposed to the corresponding positions of local maxima.
A comparison of the different features, for the various options of injection, can then be achieved and it will be discussed below.

\section{Appearance of wisps in the different scenarios and multi-band analysis}
\label{sec:res1}
\subsection{Radio and X-ray wisps}

Let us start our analysis from the scenario in which particle injection is uniform along the shock front [case (1)]. This is the simplest assumption, though probably not the best physically motivated one, and it is also the assumption that has always been adopted so far in the modelling of the Crab Nebula emission \citep[see, however][]{Porth:2014}.
The natural expectation in this case is that the wisps are largely coincident at all frequencies, with differences only due to the effects of radiation losses. These might lead to the suppression, at high energies, of features that are observed at lower energy at large distance from the shock along the streamlines. Therefore differences are expected to become more apparent in the outer part of the nebula, being longer the path taken by particles to get there.

\begin{figure}
\hspace{-5mm}
 \includegraphics[scale=.42]{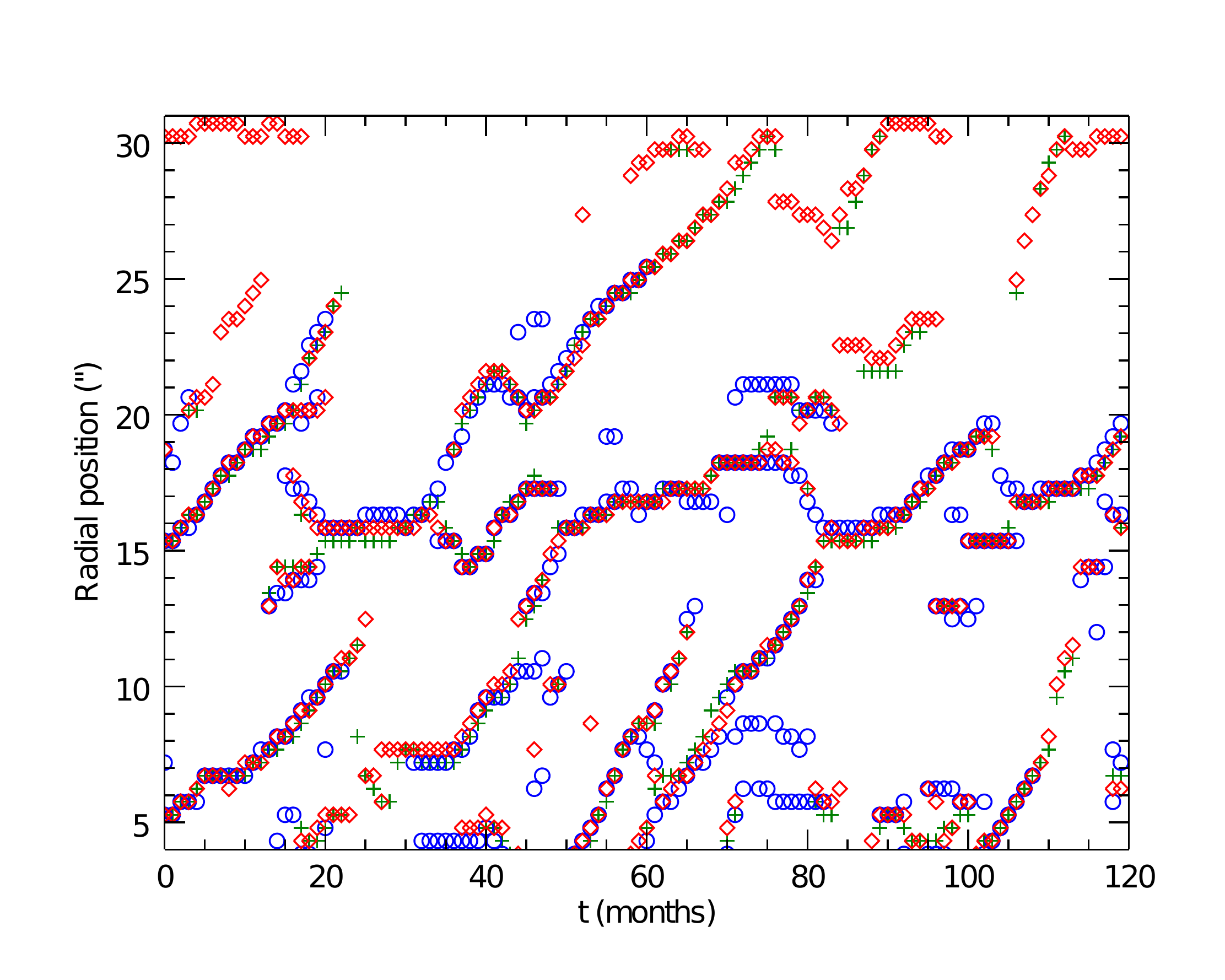}
 \caption{Radial position of local intensity maxima (in arcseconds) as a function of time (in months) referring to case (1) (i.e. uniform injection). Red diamonds are for radio emission, green crosses for optical and blue circles for the X-rays.}
  \label{fig:isoprof}
\end{figure}

In Fig.~\ref{fig:isoprof} we show the radio, optical, and X-ray wisps positions in case (1). In agreement with expectations, the emission peaks are mostly coincident at all frequencies and at all times, with the main differences appearing at distances larger than 20'' from the pulsar, where one can still see the outward motion of bright features at radio and optical frequencies, but nothing in the X-rays: this is a clear sign that losses have reduced the density of high energy particles at these places so strongly that their emission has fallen below the imposed cut-off. 
In general the brightness maxima show a motion that is mostly directed outward and has a periodicity between 1 and 2 years: this is somewhat longer, on average, than seen in real data. We checked that the discrepancy is reduced when increasing the resolution of our simulations, and chose a resolution that would lead to a wisp frequency in reasonable agreement with observations, while still keeping the computational cost tolerable. Fig.~\ref{fig:isoprof} makes it clear that, in order for wisps in the inner region not to be coincident at the different wavelengths, particles responsible for the emission in different bands must be accelerated in different sites.

A short comment on the behaviour of wisps at radio wavelengths is in order. The emission map behind the profiles in Fig.~\ref{fig:isoprof} is obtained by assuming radio particles being part of the pulsar outflow and injected uniformly along the shock surface. However, in \citet{Olmi:2014} we showed that there are no apparent differences between maps computed in a scenario where radio particles are part of the outflow or one in which they are taken as homogeneously distributed throughout the nebula at all times. We confirm that uniform injection limited to the shock surface and uniform injection in the whole nebula lead to the same wisp behaviour.

Here we have considered both scenarios once again and confirm our previous findings.

\begin{figure*}
\vspace{5mm}
 \includegraphics[scale=1.]{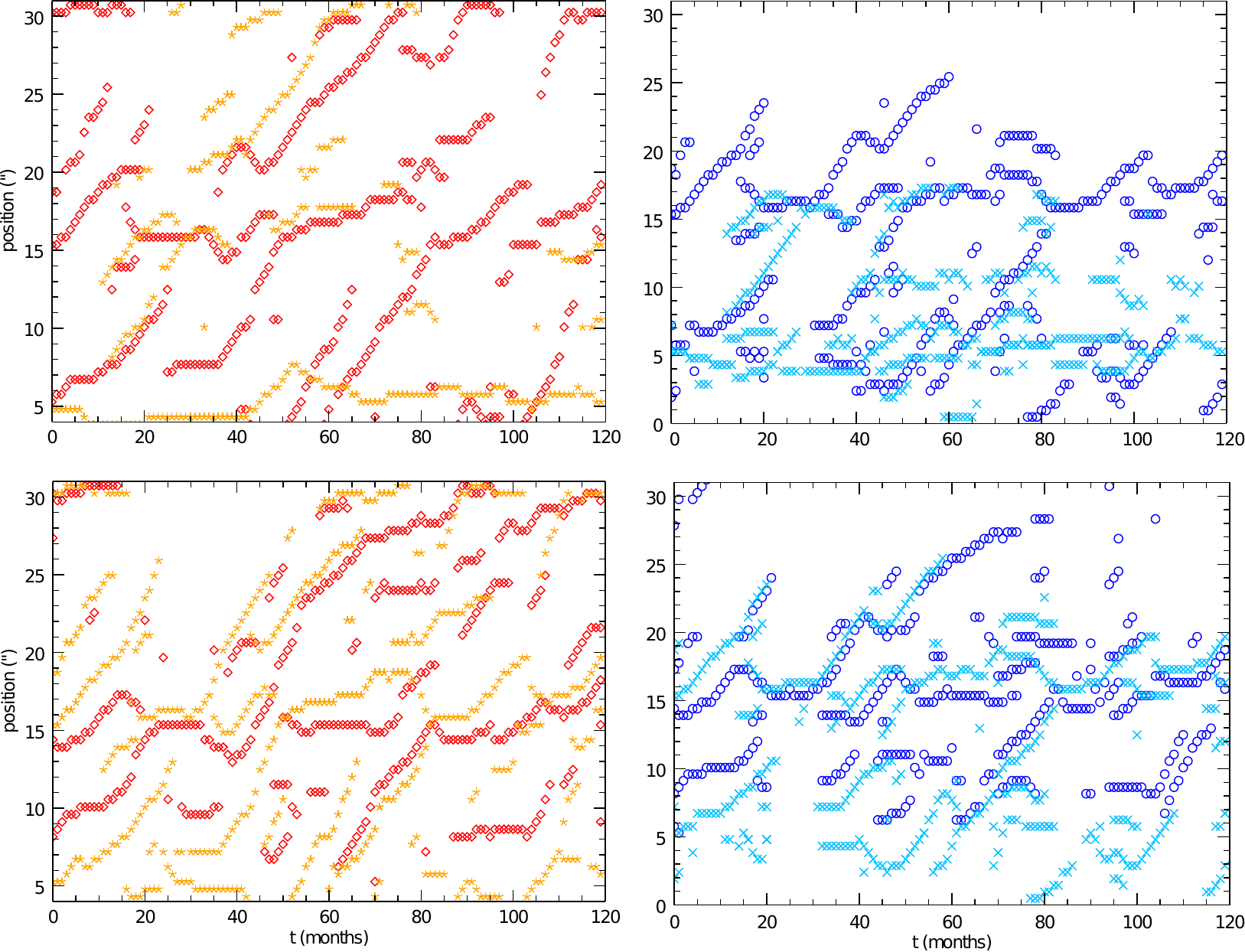}
 \caption{Position of the brightest features as a function of time at radio and X-ray wavelengths for different injection scenarios. Radio emission is on the left and X-ray emission on the right. The top row assumes a narrow polar cone [case (2)], while the bottom one is for a narrow equatorial belt [case(3)]. Injection at complementary angles is also shown in each panel. Top left panel: red diamonds are for radio emitting particles injected within an angle $\theta \in[20\degr,70\degr]$, while orange asterisks are for injection in the complementary angular sector $\theta \in[0\degr,20\degr]$. Bottom left panel: red diamonds are for injection of radio particles within $\theta \in[70\degr,90\degr]$, orange asterisks for $\theta \in[0\degr,70\degr]$. Top right panel: blue circles are for X-ray emitting particles injected within $\theta \in[20\degr,70\degr]$, while light blue crosses are for injection in the complementary angular sector $\theta \in$[$0\degr,20\degr$]. Bottom right panel: blue circles are for injection of X-ray particles within $\theta \in[70\degr,90\degr]$, light blue crosses for $\theta \in[0\degr,70\degr]$.
}
\label{fig:mixinj}
\end{figure*}

In Fig.~\ref{fig:mixinj} we show how the appearance of wisps changes at radio and X-ray frequencies when different injection sites are considered. 

First thing that is apparent from the figure is the expected behaviour that variability at radio wavelengths always extends to larger distances from the pulsar than in the X-rays: this is simply an effect of synchrotron burn-off. 

The most noticeable feature which comes out from both upper panels is the fact that for particle injection within a narrow cone around the polar axis there are basically no wisps, both at radio and at X-ray frequencies. The brightest features are more or less stationary, and the outward motion typical of wisps is rarely observed. On the contrary, when injection in a wide equatorial belt is considered, wisps appear, and they are found to be identical to those seen in the case of uniform injection (Fig.~\ref{fig:isoprof}). 

When the particles are injected according to the scenario named as case (3), the properties of observed wisps are qualitatively reproduced within both injection modes. The main difference between the polar and equatorial injection is now in the lack, in the latter scenario, of structures at distances from the pulsar shorter than about 6''. 

From comparison between the top and bottom panels it appears that many short distance features are associated with particles accelerated at intermediate latitudes along the shock front. This appears as an important point when trying to discriminate between different acceleration scenarios. In fact, in the work by \citet{Schweizer:2013} the region between 4'' and 6'' is found to be rather featureless. Within the scenarios we have considered, the only one in which prominent X-ray peaks are not seen in that range of distances is one in which X-ray emitting particles are injected in a narrow equatorial belt, that is case (3), shown in the bottom right panel of Fig.~\ref{fig:mixinj}.

\subsection{Optical wisps}

\begin{figure*}
\centering
\includegraphics[scale=.41]{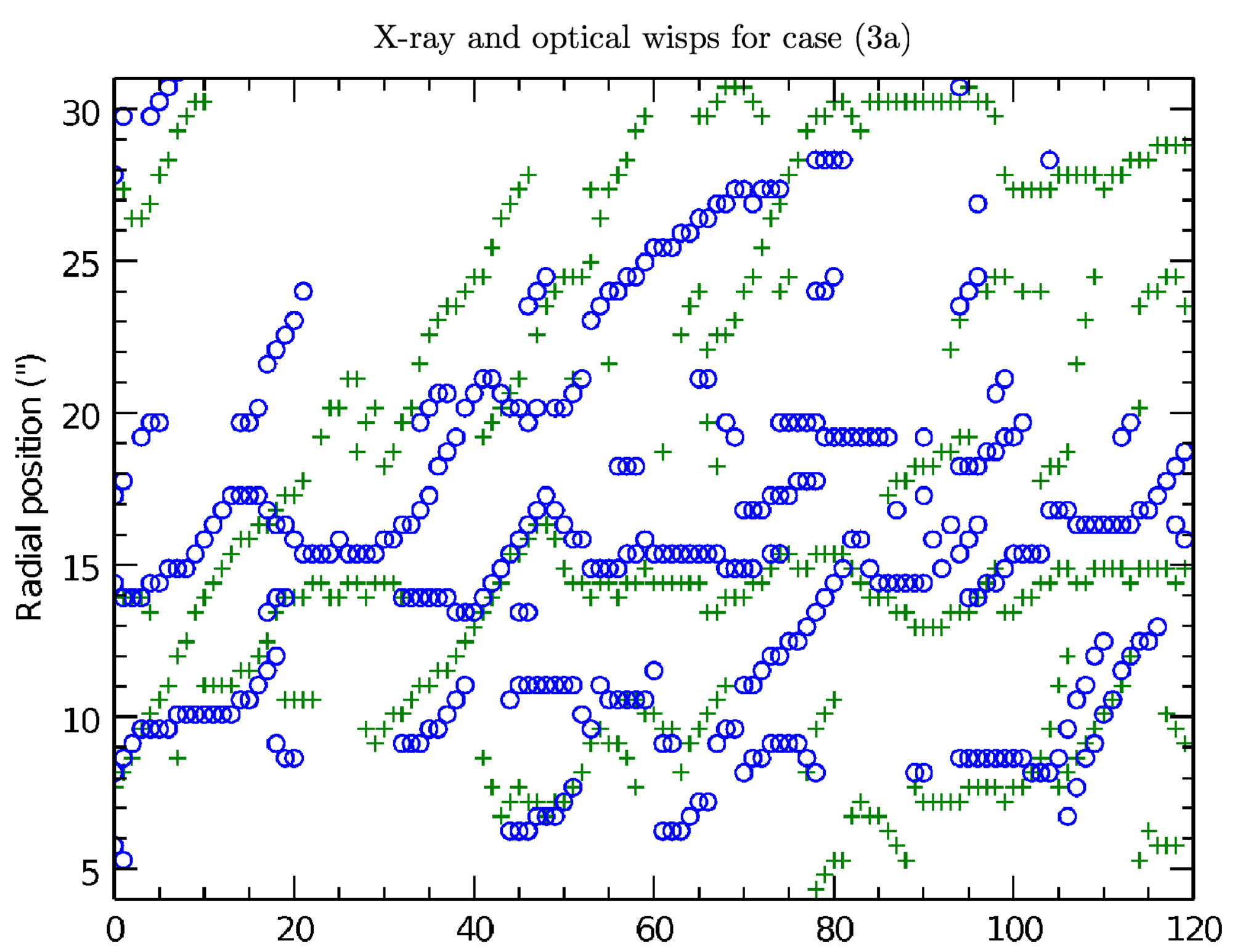}
\includegraphics[scale=.41]{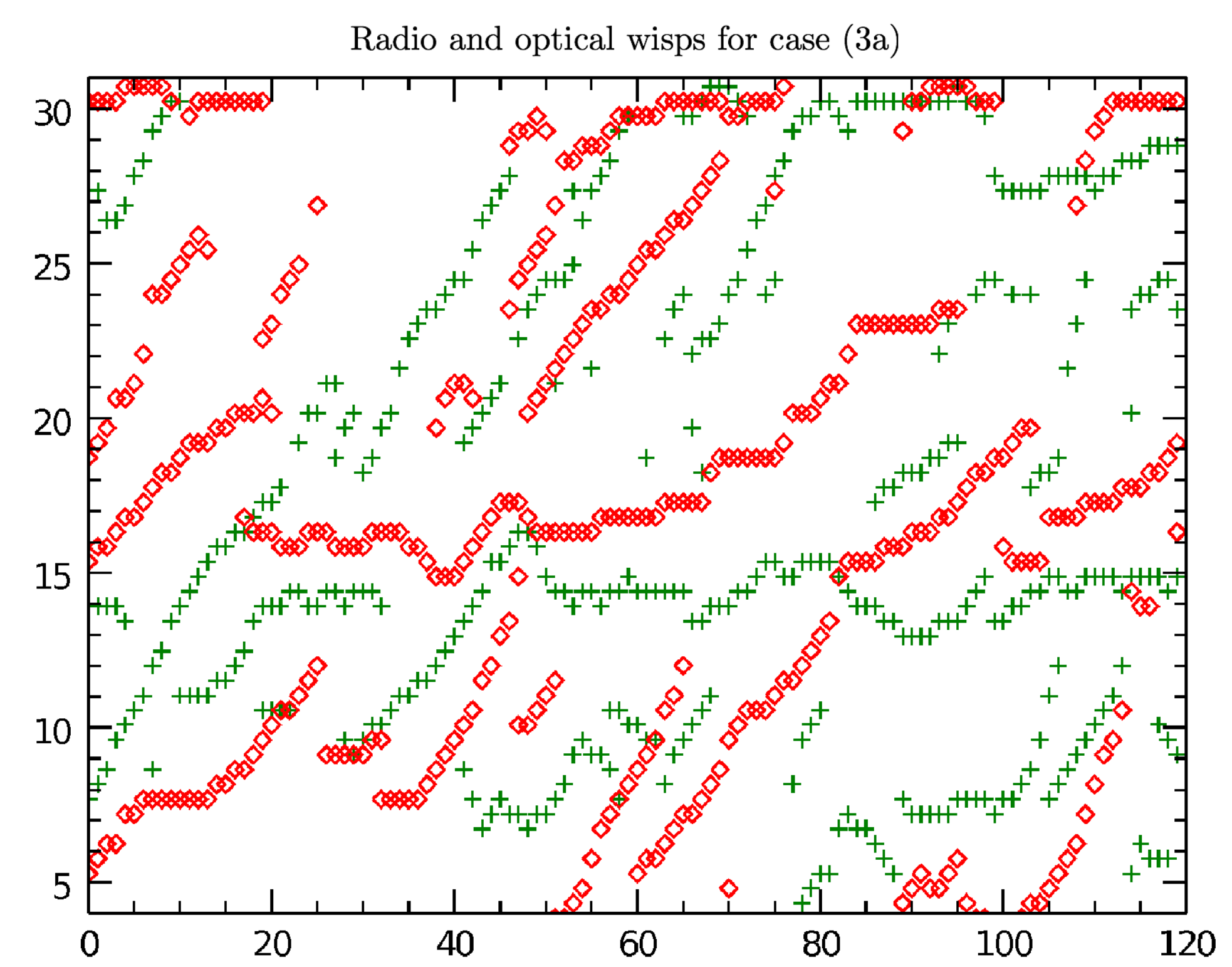}\\
\includegraphics[scale=.41]{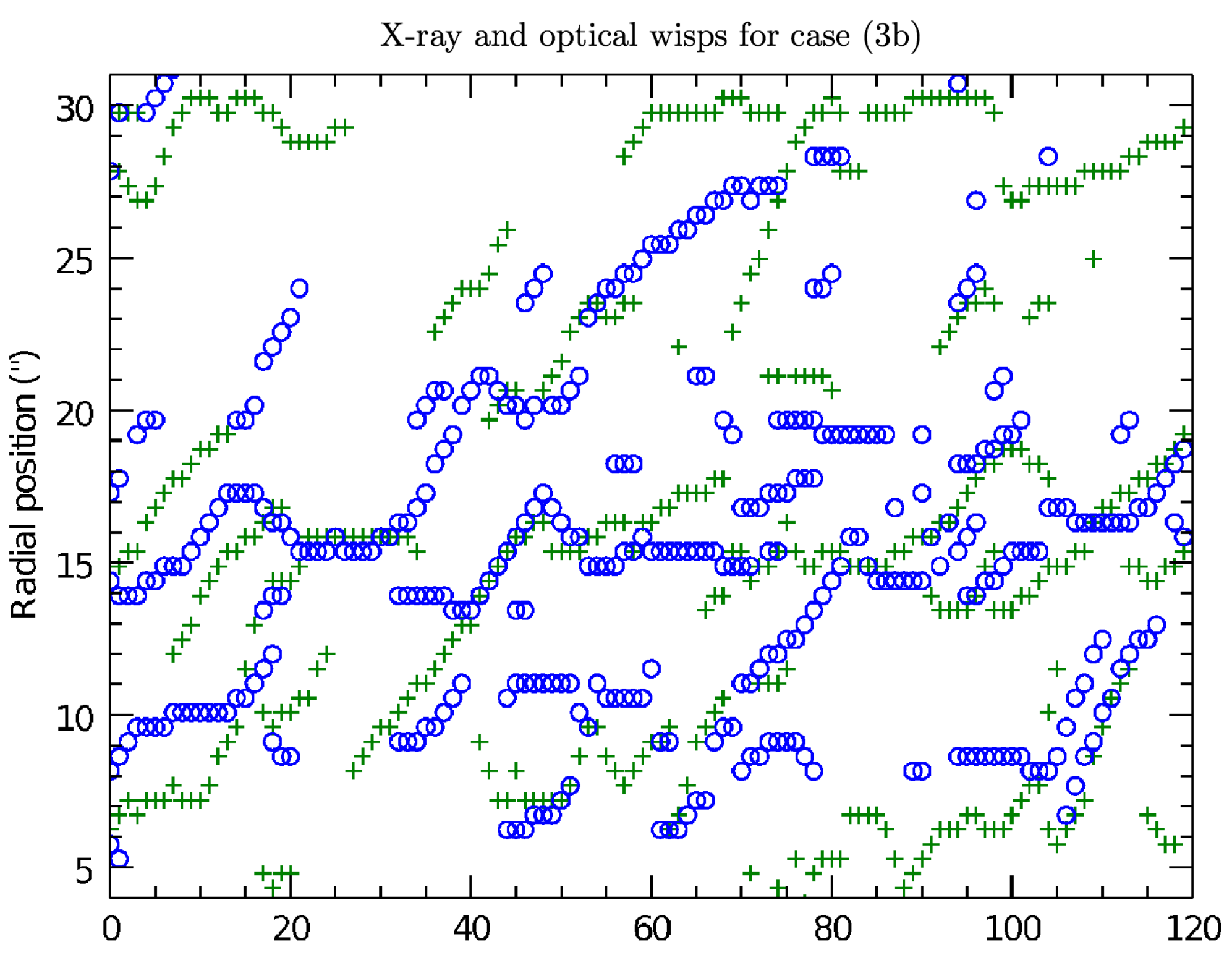}
\includegraphics[scale=.41]{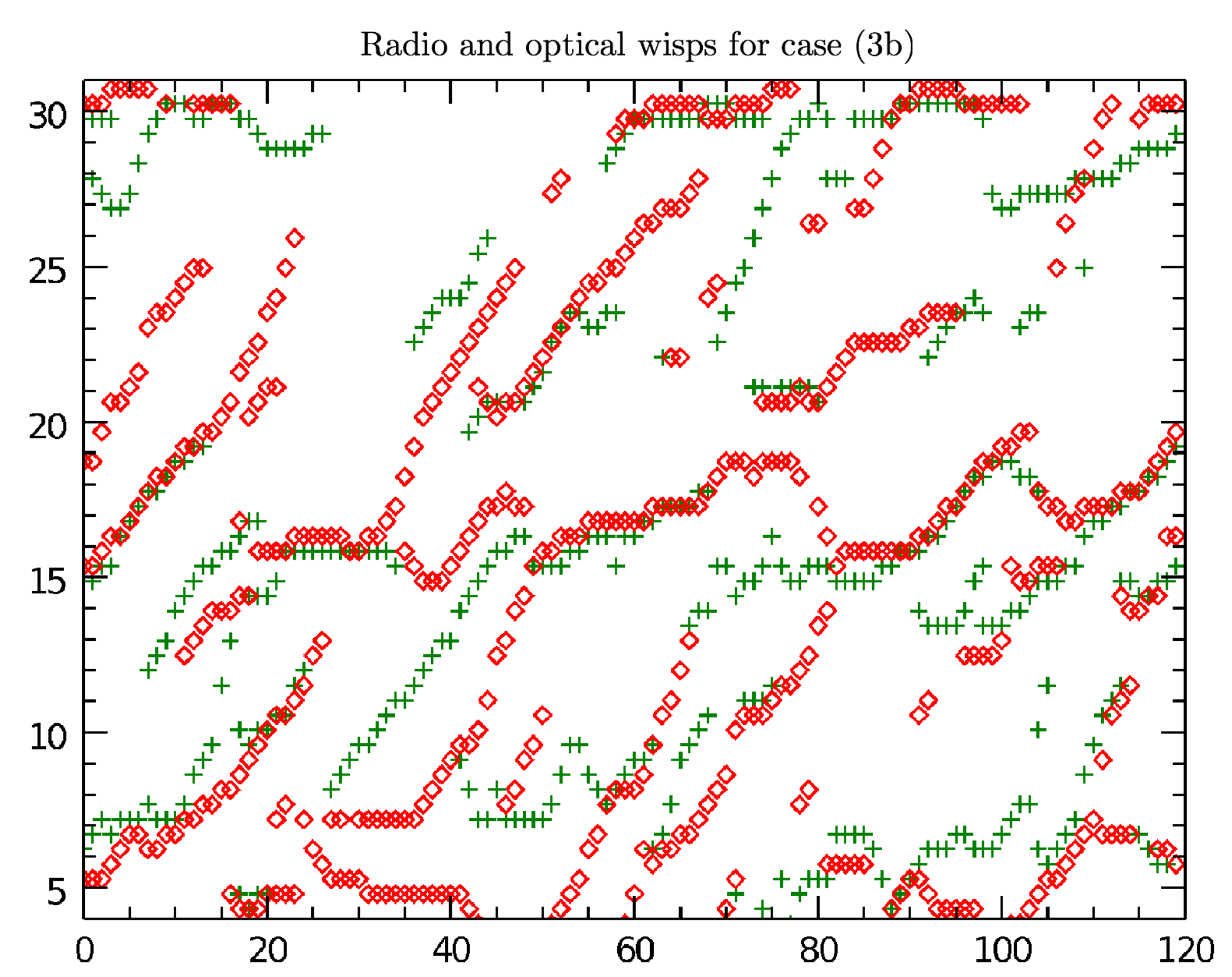}\\
\includegraphics[scale=.41]{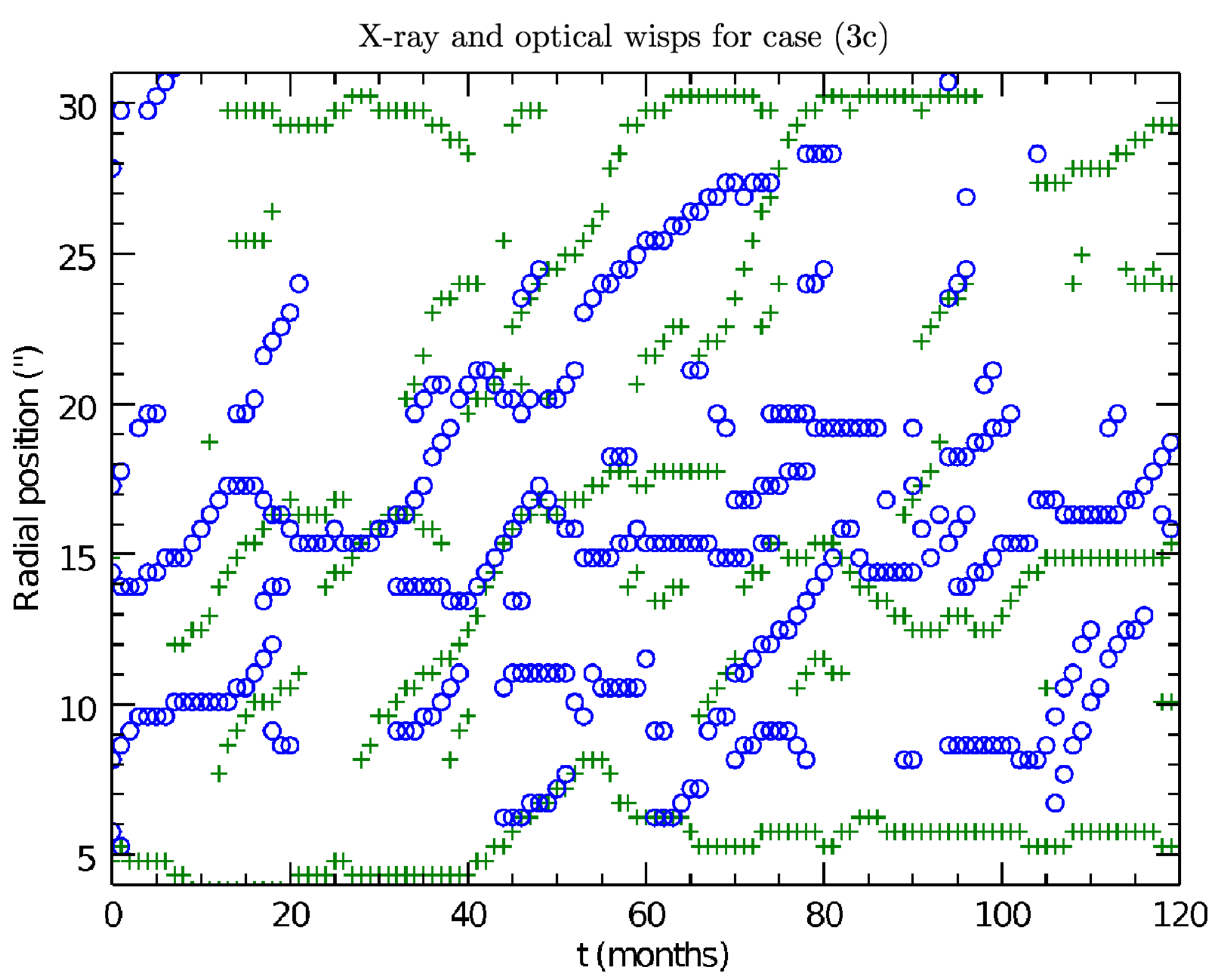}
\includegraphics[scale=.41]{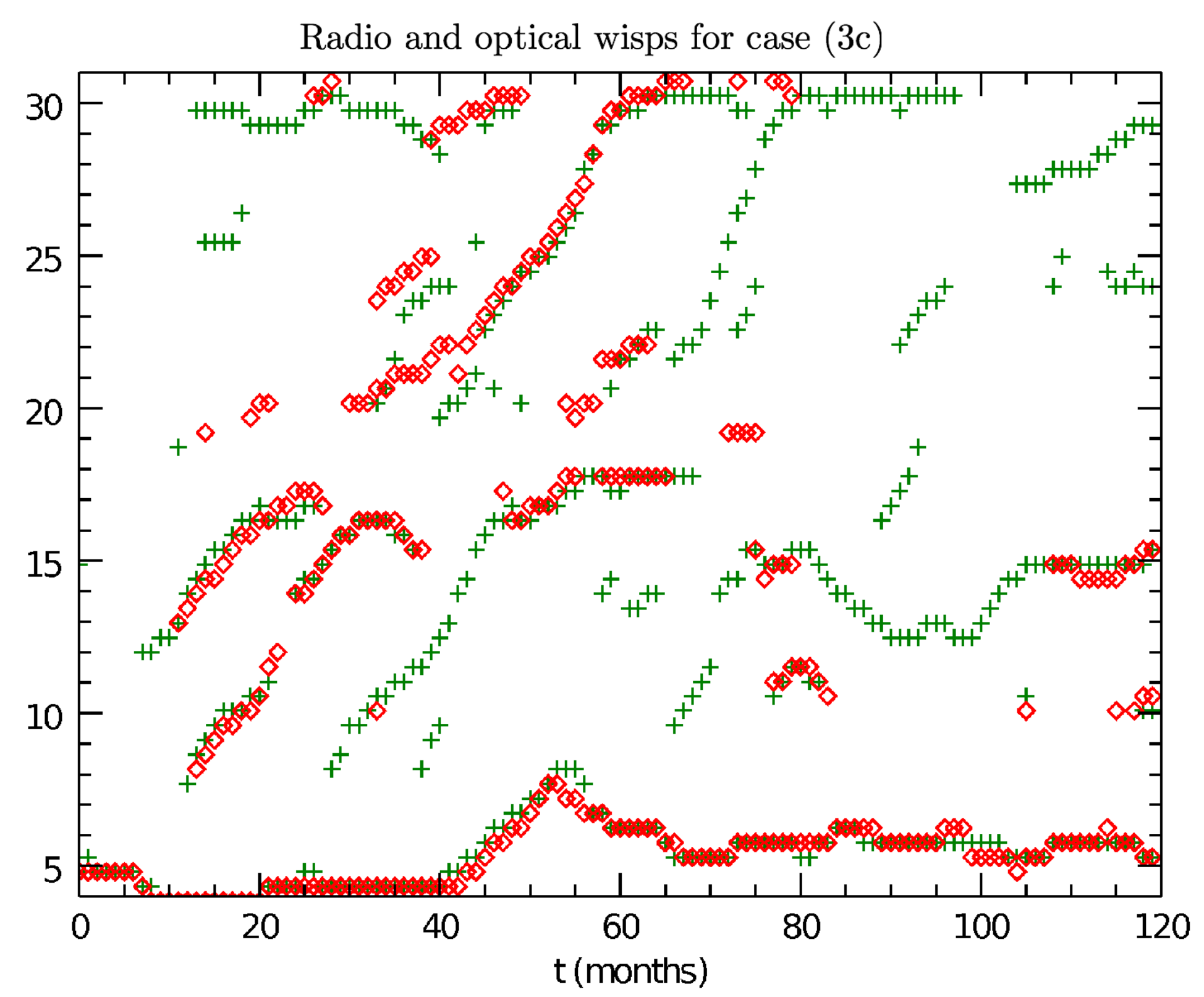}\\
 \caption{Radial distance from the pulsar of intensity peaks, as a function of time. On the left panels optical (green crosses) and X-ray (blu circles) emission profiles are compared, whereas the case of optical (green crosses) and radio (red diamonds) emission is displayed on the right panels. Here we only consider case (3) with X-ray particles injected in a narrow region around the equator $\theta \in[70\degr,90\degr]$, while radio particles are either injected uniformly (top row), in the complementary angular sector $\theta \in[0\degr,70\degr]$ (middle row), or in a narrow polar cone $\theta \in[0\degr,20\degr]$ (bottom row).} 
  \label{fig:plotprofOXR}
\end{figure*}

For the reason just discussed, in the following we focus on a scenario in which X-ray emitting particles are injected in the equatorial belt of case (3), and try to put constraints on the injection of optical emitting particles. As mentioned above, in order for optical wisps not to be coincident with X-ray ones, optical emission has to be associated with particles that have a different injection site. This condition can be satisfied if optical emission is contributed by particles that only partly belong to the high energy population, with the rest being contributed by the low-energy (radio emitting) population: in this case, differences between the X-ray and optical wisps will arise as soon as a distinct injection site for the low energy population is assumed. 

Granted that the high energy particles are injected in a narrow belt around the equator, there are three obvious choices for the injection of lower energy particles within the scenarios we have considered: (a) uniform injection; (b) injection in the complementary angular sector; (c) injection in a narrow polar cone. We show the wisps motion resulting from these three different models in the various panels of Fig.~\ref{fig:plotprofOXR}, where panels on the left refer to the comparison between optical and X-ray emission, while panels on the right side show the optical and radio cases.

We can clearly see that in all cases optical wisps are neither coincident with X-ray ones most of the time, nor to radio ones, for all our different models of injection. They typically appear at distances closer to the pulsar than the X-ray ones and extend further out. They can either lead or follow the X-ray ones. Intersections such as those observed by \citet{Schweizer:2013}, in spite of the limited period of observation, seem more rare in all cases. The cases with uniform injection of the radio emitting particles or injection in a wide polar cone, lead to results that are basically undistinguishable. While a noticeable difference in the case of injection in a narrow polar cone is the appearance of a stationary bright feature at a distance of about 5~\arcsec from the pulsar, seen in optical but not in the X-rays. 

A similar behaviour is found when comparing optical and radio emission. As for the case of optical and X-ray emission, we find that the kind of variability resulting from a scenario in which low energy particles are injected uniformly or within a wide polar cone is very similar. In the case of injection with a narrow polar cone, low energy particles give rise to the same bright quasi-steady feature discussed above,  both at radio and optical frequencies.

\section{Characterisation of wisps}
\label{sec:res2}

\begin{figure*}
 \includegraphics[scale=.35]{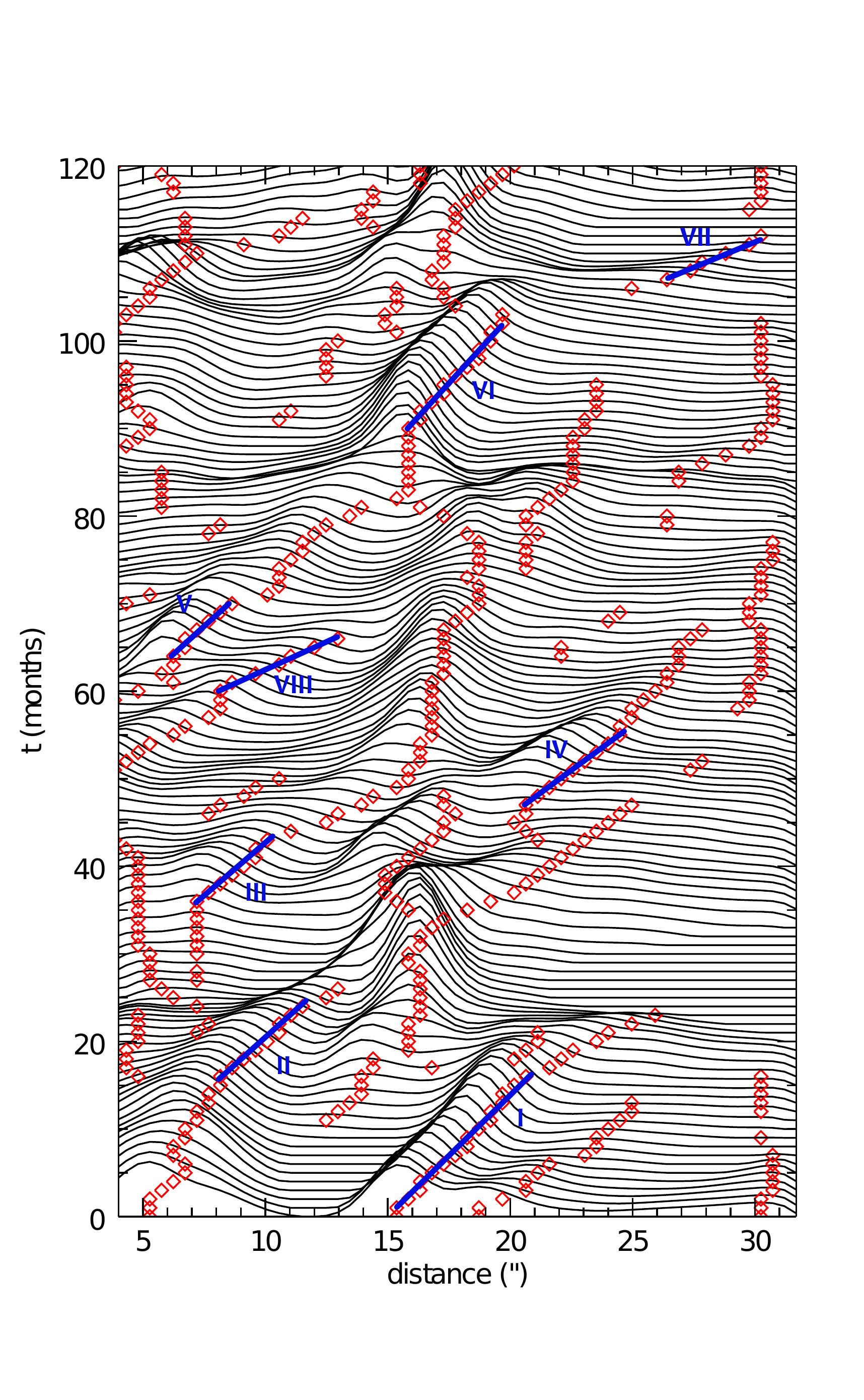}
 \hspace{-5mm}
 \includegraphics[scale=.35]{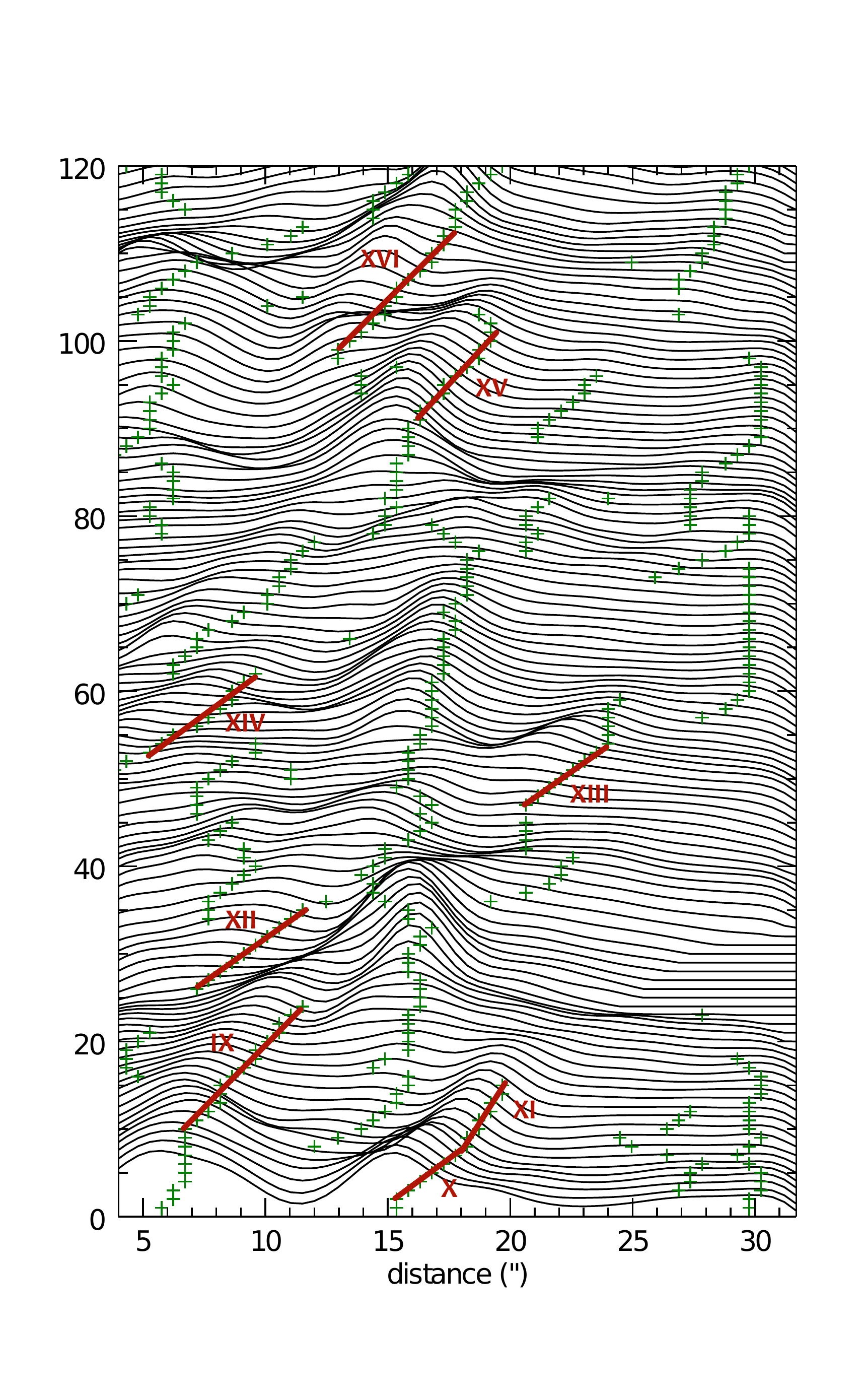}
  \hspace{-5mm}
  \includegraphics[scale=.35]{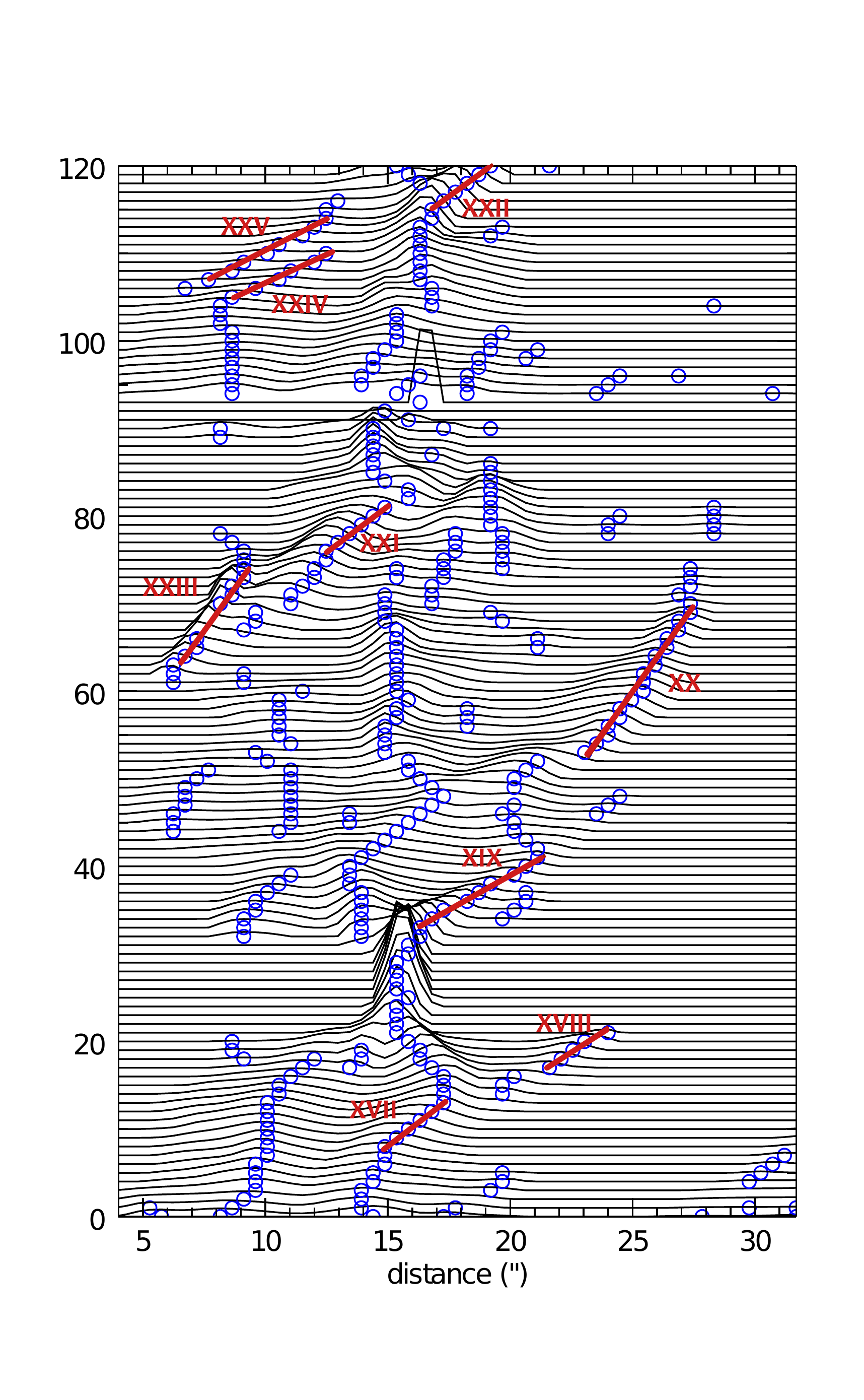}
 \caption{Stack plots with radial profiles of intensity radiation in the radio band (left panel), optical band (middle panel) and X-ray band (right panel). The results refer to case (3b), with injection of X-ray particles in the narrow equatorial belt $\theta \in[70\degr,90\degr]$ and of radio particles in the complementary wide polar region $\theta \in[0\degr,70\degr]$. Optical emission is contributed by both populations as discussed in the text. Lines and Roman numerals are used to highlight the prominent structures, and are recalled in the following for the wisp velocity evaluation (see Tab.~\ref{tabV1}).}
  \label{fig:VelOX}
\end{figure*}

Moving from qualitative to quantitative comparison with observations, the simplest analysis to be done is concerned with the velocity of the wisps in different wavebands. This was estimated by \citet{Schweizer:2013} based on their data, and is easy to evaluate from our simulations. To this aim, in Fig.~\ref{fig:VelOX} we plot the radial emissivity profiles at  radio, optical and X-ray frequencies as a function of time. The aim of this figure is that of highlighting the relative importance of the various brightness peaks. Indeed, for the velocity determination we only use the most prominent peaks. In each plot it is possible to follow the time and spatial evolution of a wisp by identifying what appears to be the same peak at a different time and position.

\begin{table*}
\centering
\begin{tabular}{cccccccccccc}
\hline
  \multicolumn{4}{c}{Radio} &  \multicolumn{4}{c}{Optical} &  \multicolumn{4}{c}{X-ray} \\
  Wisp \# & $\varv_\mathrm{app}$ (\arcsec/d) & $\varv_\mathrm{app}/c$ & $\varv/c$ &  Wisp \# & $\varv_\mathrm{app}$ (\arcsec/d) & $\varv_\mathrm{app}/c$ & $\varv/c$ & Wisp \# & $\varv_\mathrm{app}$ (\arcsec/d) & $\varv_\mathrm{app}/c$ & $\varv/c$  \\
 \hline
I       &  0.012   &  0.14  &  0.16 & IX     & 0.011   & 0.13   &  0.15  & XVII     & 0.016    &  0.18  & 0.21  \\
II      &  0.014   &  0.16  &  0.18 & X      & 0.016   & 0.18   &  0.21  & XVIII    &  0.018   &  0.21  & 0.24  \\
III     &  0.016   &  0.18  &  0.21 & XI     & 0.007   & 0.08   &  0.09  & XIX      &  0.020   &  0.23  & 0.27 \\
 IV   &  0.015   &  0.17  &  0.20 & XII    & 0.016   & 0.18    &  0.21  & XX      &  0.009   &  0.10   & 0.12 \\ 
V     &  0.013   &  0.15  &  0.17 & XIII   & 0.015   & 0.17   &  0.20   & XXI     &   0.013  &  0.15   & 0.17\\
VI    &  0.010   &  0.11  &  0.13 & XIV   &  0.016 & 0.18   &  0.21  & XXII      &  0.018    &  0.21  & 0.24  \\
VII   &  0.027   &  0.31  &  0.36 & XV    & 0.011  & 0.13    &  0.15  & XXIII    &  0.009    &  0.10  & 0.12  \\
VIII  &  0.029   &  0.33  &  0.38  & XVI   &  0.011 & 0.13   &  0.15  & XXIV   &  0.026   &  0.30   & 0.35 \\
        &              &             &            &           &            &             &            &XXV      & 0.023    &  0.26  & 0.30   \\ 
 \hline
\end{tabular}
 \caption{Measured apparent velocity $\varv_\mathrm{app}$ (in \arcsec/d) and inferred real velocity $\varv/c$ for case (3), in the radio, optical and X-ray bands. Roman numerals refer to the prominent wisps shown in Fig.~\ref{fig:VelOX}.}
\label{tabV1}
\end{table*}

By connecting all the peaks related to a given wisp, one obtains a line that traces the evolution of that wisp (as reported in Fig.~\ref{fig:VelOX}). The apparent velocity is then given by the slope of that line, and can be directly compared to that inferred from real observations of wisps. Obviously the derived velocities $\varv_\mathrm{app}$ are those on the plane of the sky, the real ones $\varv$ can then be obtained by deprojection, making use of the known CN distance ($\simeq 2$~kpc) and its inclination angle ($\simeq 60\degr$) \citep[see][]{Weisskopf:2012}. Here we focus only on case (3), that is the one that better reproduces, at least qualitatively, the observed properties of the X-ray wisps. The inferred velocities $\varv_\mathrm{app}$ and $\varv$ are given in Tab.~\ref{tabV1}. 

The resulting velocities show almost the same mean values for the three families: $\varv_\mathrm{app,R}=(0.017 \pm 0.006)\,\arcsec/\rm{d}$, $ \varv_\mathrm{app,O}=(0.013 \pm 0.003)\,\arcsec/\rm{d}$ and $\varv_\mathrm{app,X}=(0.017 \pm 0.007)\,\arcsec/\rm{d}$, that deprojected and in units of $c$ become $\varv_\mathrm{R}\simeq \varv_\mathrm{X}\simeq 0.2 c$, $\varv_\mathrm{O}\simeq 0.15 c$. 
The deprojected velocities range is $0.08 c$ to $0.38 c$, and it is in good agreement with those extrapolated by the observed data: in the work by \citet{Schweizer:2013} optical and X-ray wisps appear to have the same outward speeds, ranging from $0.16 c$ to $0.44 c$. 

The projected speed of optical wisps was also estimated by \citet{Bietenholz:2004}, who found it to be $\lesssim 0.3 c$. 
Based on the comparison of the optical and radio images, the authors also argue that radio wisps appear to have a somewhat lower speed than the optical ones. One could expect that the same differences should be found between X-ray and radio wisps. Unfortunately the very small differences in speed that we find in our analysis do not allow us to draw any firm conclusions on this trend. 

Moreover \citet{Bietenholz:2004} find that radio wisps cover a larger area of the nebula, compared to the optical ones, and typical velocities decrease with the distance from the pulsar. 
As expected, the simulated X-ray wisps vanish beyond $30\arcsec$ (1~ly). On the contrary, many radio wisps are still visible up to $60\arcsec$. A slight decrease in velocity appears around $50\arcsec$, where the wisps show speeds compatible with the lower ones in Tab.~\ref{tabV1}.

\begin{figure}
 \includegraphics[scale=.58]{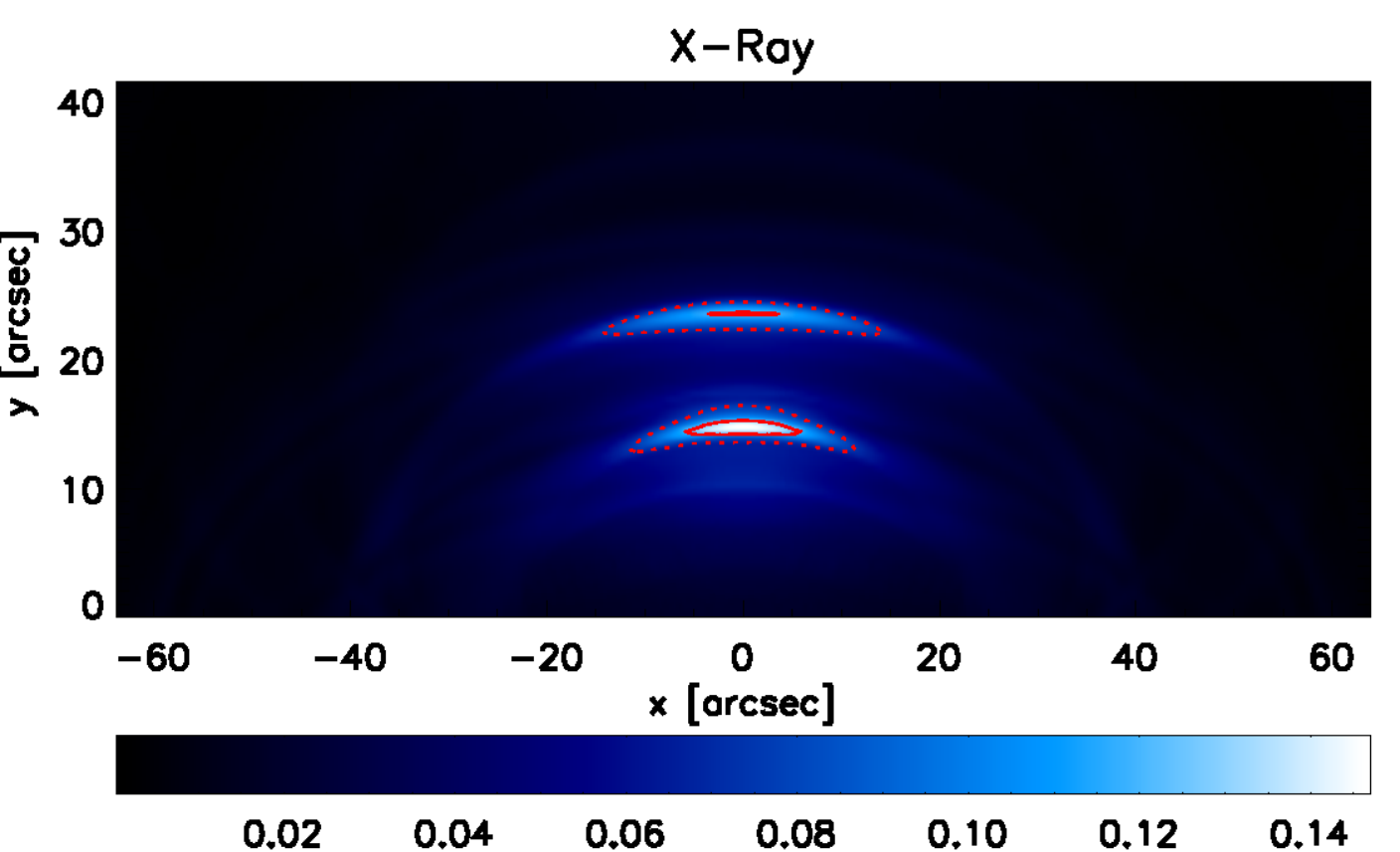}
 \includegraphics[scale=.58]{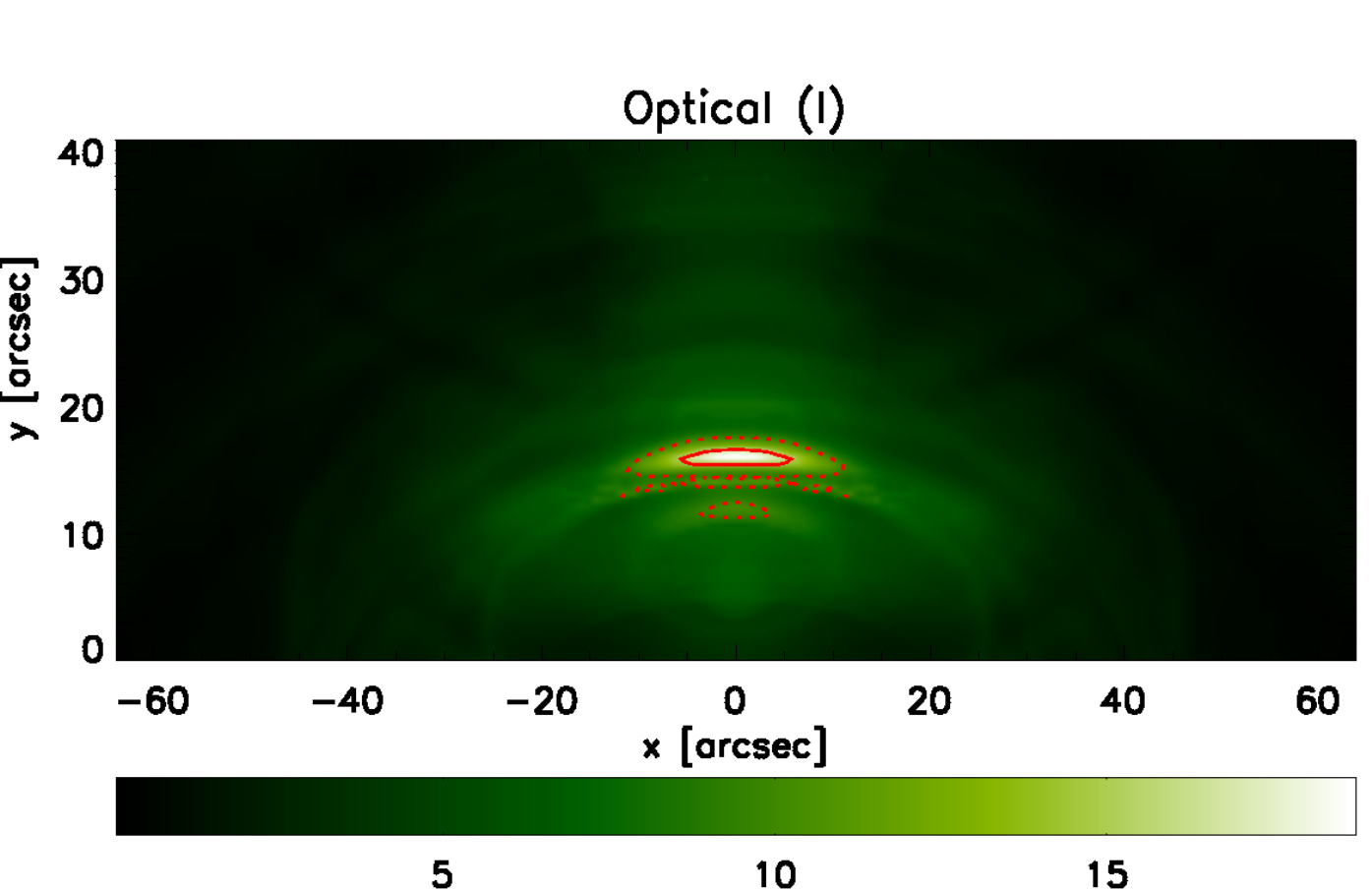}
 \caption{Wisps in the inner nebula with contour lines at $0.5 I_\mathrm{max}$ (dashed) and $0.8 I_\mathrm{max}$ (solid). X-ray in the top panel and optical in the bottom one, both in mJy/arcsec$^2$.}
  \label{fig:ellipse1}
\end{figure}

Finally we focus on the azimuthal profile of wisps, again following the analysis by \citet{Schweizer:2013}. They find an angular size of the intensity profiles for the observed wisps of $\sim 35 \degr$ in the X-rays and of $\sim 15 \degr$ in the optical band. 
In our analysis we set up the same cutoff in intensity as described previously in \S \ref{sec:data_analysis}, and contour lines are drawn for $0.5 I_\mathrm{max}$ and $0.8 I_\mathrm{max}$ (two examples are shown in Fig.\ref{fig:ellipse1}). 
In order to compare our results with real data, only those structures nearest to the pulsar are considered, in the upper hemisphere. Each wisp is deprojected from the plane of the sky and then fitted with a suitable ellipse, and the angular extent is determined as the opening angle of the ellipse arc coincident with the wisp. 
This analysis is repeated on each one of our maps. For direct comparison with \citet{Schweizer:2013} we then average all of our results obtaining $(39 \pm 12)\degr$ for the size of the X-ray wisps and $(47\pm 16)\degr$ for the optical ones.
The X-ray value is in perfect agreement with observations, while the optical one is overestimated. 

This was already pointed out and discussed by \citet{Schweizer:2013} using a toy model.

\section{SUMMARY AND CONCLUSIONS}\label{sec:concl}
In this article we have tried to put constraints on the sites of particle acceleration in the Crab Nebula. Our aim was to use the information deriving from studies of the nebular variability at different frequencies. The motivation for this work came from observations of the nebula at various times and at different frequencies, which have shown that wisps, namely the brightest moving features, are not coincident at the different wavelengths: wisps are distinct in radio, in optical and in X-rays. In a MHD description of the flow, where wisps only arise as a result of Doppler boosting and magnetic field enhancement, this cannot be explained, unless particles responsible for emission at different frequencies have different acceleration sites. We studied this possibility, by performing MHD simulations that assumed particle acceleration at the shock to be non-uniform. In particular we divided the shock front in complementary regions: an equatorial band and a polar cone, of varying angular extent.

We found that the properties of X-ray wisps are best reproduced if injection in a narrow equatorial band is considered. The most important piece of evidence pointing towards this conclusion is the absence of bright X-ray features very close to the pulsar. The observations by \citet{Schweizer:2013} show that X-ray brightness peaks are never observed at distances from the pulsar shorter than $\sim 6$\arcsec. 
In our simulations we find that the lack of X-ray peaks very close to the pulsar is only obtained when injection of the particles responsible for the emission is confined to a narrow angular sector around the equator. This is easily understood by looking at the flow structure shown in Fig.~\ref{fig:TS-zones}. While it is clear that in the equatorial region there are no vortices and the flow is everywhere directed outwards, in the region between the lines marked as (2) and (3) vortices are seen almost at all times, though varying in position, and these are the structures responsible for the innermost brightness enhancements. 
 
That X-ray emitting particles are most likely injected only in the vicinities of the equator was also suggested by the work of \citet{Porth:2014}. These authors considered two different models of injection of the X-ray emitting particles within their 3D MHD simulations: uniform throughout the shock front and equatorial only injection. Due to the modest spatial resolution of their simulation, enforced by the full 3D setup, they could only see two wisps, but were still able to conclude that equatorial injection works better at high energies, based on comparison, in the two cases, of the brightness contrast of prominent X-ray features as the jet and the anvil.  

One might wonder whether and how the X-ray knot fits in this scheme, since this especially bright feature has usually been associated with the flow at relatively high latitude. While we focus here on the wrong hemisphere as far as the knot is concerned, and we would not be able to see it in any case, the question is certainly relevant. No knot is in fact obtained if X-ray emitting particles are only accelerated in a narrow equatorial sector. On the other hand, uniform injection of high energy particles along the shock front leads to a simulated knot much brighter than observed. A possibility is that some high energy particles are also accelerated at higher latitudes but with much lower efficiency, similar to what happens in the model by \citet{Porth:2014}, who indeed find a knot with a reasonable luminosity contrast.

The fact that different simulations, with different dimensionality, criticalities, and a focus on various observed properties, reach the same conclusion is reassuring, and might suggest that we are really pinning down the acceleration site of the highest energy particles in Crab.

Moreover we also looked at the spectral properties of the optical wisps, by computing their spectral index as in \citet{Del-Zanna:2006}, using the specific intensities at $\nu_\mathrm{I}=3.75 \times 10^{14}$ Hz (I band) and $\nu_\mathrm{R}=4.6 \times 10^{14}$ Hz (R band). We measured the spectral index for each injection scenario, focusing on the most prominent wisp structures. 
We found that when X-ray particles are injected into the equatorial belt the spectral index is roughly uniform and its value is $\sim 0.5 \div 0.6$, in good agreement with observations \citep{Veron-Cetty:1993}, regardless
of the modality for radio particles injection. On the contrary, when X-ray particles are injected in the polar zone, the spectral index behaviour is not compatible with observations. 

Moving to a quantitative comparison of our results with observations, one thing that is confirmed by this study is that almost all the properties of the wisps can be explained purely as a result of the MHD flow structure. 

Our model reproduces the observed velocities at both optical and X-ray frequencies. In the radio band, the only available measurement to our knowledge is that by \citet{Bietenholz:2004}, who measure velocities of the radio wisps of order 0.3 c (see also \citet{Bietenholz:2015} and \citet{Lobanov:2011} for the time variability of the wisps): this is also in excellent agreement with our findings.

In our analysis we do not find any features moving with $v \sim c$: this fact, while compatible with observations, might seem somewhat surprising. In our simulations, the bulk Lorentz factor beyond the termination shock is generally below 2, but this might represent an underestimate of real velocities in the nebula, because the highest Lorentz factors are expected to be found in the polar region which in reality will be much more variable than in our axisymmetric simulations. On the other hand, the lack of observational evidence for such fast motion could be due to effective deboosting of the emitting features.

Finally, we have compared our simulation results with observations of the angular profiles of emissivity of the wisps. We found excellent agreement at X-ray frequencies, but could not account for optical observations. This problem and possible solutions to it were already discussed by \citet{Schweizer:2013}.

Moving to lower energies, putting constraints on particle injection based on radio emission is much more complicated, as we already partly discussed in \citet{Olmi:2014}. The only scenario that radio observations directly disfavour is one in which the corresponding particles are injected in a narrow polar cone: in this case variability at radio wavelengths is much reduced, radio wisps are rare and the brightness peak in this waveband is constantly found at a distance from the pulsar of order 6\arcsec. 

Uniform injection and injection in a wide equatorial or polar sector lead to qualitatively very similar results from the point of view of radio emission, as can be seen from comparison of the red diamonds in Fig.~\ref{fig:isoprof} with the red diamonds in the top left panel of Fig.~\ref{fig:mixinj} and the orange asterisks in the lower left panel of the same figure. The main observed properties in this waveband are well reproduced in all these cases and even assuming a uniform distribution of emitting particles throughout the nebula. 

However optical emission can be used to put additional constraints. When optical emission is compared in the different scenarios, a slight preference for uniform injection of radio particles is suggested by the fact that this leads to brightness peaks in the optical that are in general at distances from the pulsar larger than 5\arcsec, in agreement with the findings of \citet{Schweizer:2013}. It is also worth remarking that even at optical frequencies, we find it difficult to distinguish the emission computed in the case when radio particles are uniformly distributed through the nebula and that in which they are accelerated at the termination shock (uniformly along its surface) and advected with the flow.

On the other hand, from a more theoretical point of view, the lack of a real discontinuity in the nebular emission spectrum between the radio and X-rays suggests that the acceleration of lower and higher energy particles cannot be due to completely uncorrelated phenomena. It is natural, in a sense, to think that less extreme conditions are required to accelerate particles to the energies required for radio emission, than to bring them to
the ones need to emit synchrotron X-rays. In this view, one might expect that radio particles are accelerated in a wide region, while higher energies are reached only within a portion of this region. The flat spectrum of radio emitting particles suggests that their acceleration is connected with magnetic reconnection, which is expected to occur in a region around the equator with angular width corresponding to the inclination between the pulsar magnetic and rotation axis. A possibility is that X-ray energies are then reached only in a smaller region, when field dissipation is such that the magnetization drops to low enough values to allow Fermi acceleration to occur, which would also agree with the steeper spectrum of higher energy particles. However, what is making the acceleration of lower energy particles still stays as a very deep mystery.

\footnotesize{
\bibliographystyle{mn2e}
\bibliography{olmi}
}

\label{lastpage}

\end{document}